\documentclass[aps,prxq,showpacs,superscriptaddress,tightenlines,twocolumn,lengthcheck]{revtex4-2}
\usepackage{amssymb}
\usepackage{amsmath}
\usepackage{mathtools}
\usepackage{graphicx}
\usepackage{txfonts}
\usepackage{color}

\usepackage{amsfonts}
\usepackage[colorlinks,urlcolor=blue,linkcolor=blue,citecolor=blue]{hyperref}
\begin{document}
\title{Reshaping coupled bosonic networks: A bipartite-graph framework for optimal quantum excitation transfer}

\author{Cheng Liu}
\affiliation{Key Laboratory of Low-Dimensional Quantum Structures and Quantum Control of Ministry of Education, Key Laboratory for Matter Microstructure and Function of Hunan Province, Department of Physics and Synergetic Innovation Center for Quantum Effects and Applications, Hunan Normal University, Changsha 410081, China}
\affiliation{Hunan Research Center of the Basic Discipline for Quantum Effects and Quantum Technologies, Hunan Normal University, Changsha 410081, China}
\author{Yu-Hong Liu}
\affiliation{Key Laboratory of Low-Dimensional Quantum Structures and Quantum Control of Ministry of Education, Key Laboratory for Matter Microstructure and Function of Hunan Province, Department of Physics and Synergetic Innovation Center for Quantum Effects and Applications, Hunan Normal University, Changsha 410081, China}
\affiliation{Hunan Research Center of the Basic Discipline for Quantum Effects and Quantum Technologies, Hunan Normal University, Changsha 410081, China}
\author{Le-Man Kuang}
\affiliation{Key Laboratory of Low-Dimensional Quantum Structures and Quantum Control of Ministry of Education, Key Laboratory for Matter Microstructure and Function of Hunan Province, Department of Physics and Synergetic Innovation Center for Quantum Effects and Applications, Hunan Normal University, Changsha 410081, China}
\affiliation{Hunan Research Center of the Basic Discipline for Quantum Effects and Quantum Technologies, Hunan Normal University, Changsha 410081, China}

\author{Franco Nori}
\affiliation{Quantum Computing Center, RIKEN, Wakoshi, Saitama 351-0198, Japan}
\affiliation{Department of Physics, The University of Michigan, Ann Arbor, MI, 48109-1040, USA}

\author{Jie-Qiao Liao}
\email{Contact author: jqliao@hunnu.edu.cn}
\affiliation{Key Laboratory of Low-Dimensional Quantum Structures and Quantum Control of Ministry of Education, Key Laboratory for Matter Microstructure and Function of Hunan Province, Department of Physics and Synergetic Innovation Center for Quantum Effects and Applications, Hunan Normal University, Changsha 410081, China}
\affiliation{Hunan Research Center of the Basic Discipline for Quantum Effects and Quantum Technologies, Hunan Normal University, Changsha 410081, China}
\affiliation{Institute of Interdisciplinary Studies, Hunan Normal University, Changsha, 410081, China}

\begin{abstract}
Highly efficient transfer of quantum resources including quantum excitations, states, and information on
coupled bosonic networks is an important task in quantum physics and quantum information science. Here we propose a bipartite-graph framework to characterize quantum excitation transfer in coupled bosonic networks. This is achieved by diagonalizing the intermediate subnetwork between the sender and the receiver to construct a bipartite-graph configuration, and hence this treatment can be understood from the viewpoint of network deformation. We examine the covariance matrix of the coupled bosonic networks in both the original and bipartite-graph representations. In particular, we investigate quantum excitation transfer in both the finite and infinite intermediate-normal-mode cases and show the dependence of the transfer efficiency on the network configurations and system parameters. We find the bounds of maximally transferred excitations for various network configurations and reveal the underlying physical mechanisms. We also discover that the dark-mode effect will degrade the excitation transfer efficiency. Our findings provide a new insight for the design and optimization of networks in physics, information theory, and complex system science.
\end{abstract}
\date{\today}
\maketitle

\section{\label{sec:level1}Introduction}
Coupled quantum networks~\cite{Koll2019} are typically composed of quantum nodes, which are connected by quantum couplings. The construction of coupled quantum networks is one of the most important tasks in quantum physics and quantum information science~\cite{Braunstein2005}. This is because the coupled quantum networks can not only enlarge the scale of information-processing nodes but also overcome the physical limitations of individual quantum processors and then break the spatial constraints. Currently, coupled quantum networks have been constructed in various physical platforms, such as ion traps~\cite{Kielpinski2002}, coupled cavities and waveguides~\cite{Cirac1997,Christodoulides2003,Hartmann2006,Greentree2006,Hartmann2008,Notomi2008,Alberto2010,Houck2012,Habraken2012,Zhou2013,Vermersch2017,HaoT2018,Tobias2022,Saxena2023,Yang2024,TangHao2024,Yu2025}, coupled-oscillator  networks~\cite{Ren2010,Zheng2010,Martens2013,Okamoto2013,Zhang2013,Pietras2019,Csaba2020,Zalalutdinov2006,HuangPu2013,Pietras2019,Csaba2020,Wanjura2023}, spin networks~\cite{Bose2003,Christandl2004,Christandl2005,Yung2005,Franco2008,Yao2011,Yao2013,Patil2022,Reitz2022,Xiang2024}, optomechanical networks~\cite{Heinrich2011,Ludwig2013,Aspelmeyer2014,Xuereb2014,Zhang2015,Dong2015,Peterson2017,Arnold2020,delPino2022,Slim2025}, and synthetic photonic lattices~\cite{Regensburger2012,Celi2014,Schmidt2015,Yuanluqi2019,Lustig2019,Ozawa2019,Lustig2021,Leefmans2022,Parto2023,Leefmans2024}. 

To exploit the advantages of coupled quantum networks, the implementation of highly efficient quantum resource transfer on networks becomes a significant and desired task. As a result, how to understand the physical principle determining the network function, design the network configuration and parameters, and broaden the applications of networks become critical scientific topics in this field. In recent years, many achievements have been made in the study of quantum networks, such as quantum transfer~\cite{Kielpinski2002,Bose2003,Subrahmanyam2004,Christandl2004,Albanese2004,Christandl2005,Yung2005,Feder2006,Franco2008,Wu2009,Yao2011,Yao2013,Nikolopoulos2014,Chapman2016,Kurpiers2018,Li2018,Slussarenko2019,Xiang2024,Zhou2024,Kristjansson2024,Yang2025,JinJ2026}, quantum walks~\cite{Caser1996,Alberto2010,Faccin2013,Philipp2015,Chakraborty2016,Zhiguang2019,Su2019,Karuna2021,Ming2021,Esposito2022,Lin2023,Meng2024}, entanglement distribution~\cite{Kimble2008,Stephanie2018,Cirac1997,Chin2007,Perseguers2008,Sun2016,Zhong2021}, and quantum synchronization~\cite{Vinokur2008,Lohe2010,Roulet2018,Lorenzo_2022,Nadolny2023}. 
Among these topics, efficient quantum resource transfer plays a crucial role in ensuring the reliability of quantum information processing, optimizing the utilization of distributed quantum resources, and facilitating the advancement of scalable and resilient quantum networks. It has been found that highly efficient quantum resource transfer on a static network can be realized based on preengineered couplings among nodes, without using complex dynamical control~\cite{Bose2003,Christandl2004,Albanese2004}.  These transfer protocols have mainly been developed for chain-type networks with prescribed coupling patterns. However, determining the optimal parameter conditions for preengineering couplings is highly challenging in general quantum networks due to the complexity of the  transmission paths. Furthermore, the dark-mode effect originated from destructive interference will also impede the perfect transfer of quantum resources. Therefore, optimizing the coupling configurations of networks, analyzing the influence of dark-mode effects on quantum transfer, and determining the parameter conditions for achieving highly efficient quantum resource transfer are important and desired research topics in this field. Motivated by these challenges, we introduce a bipartite-graph framework that maps general coupled bosonic networks onto a simple representation. This framework allows us to identify the roles of coupling paths and dark modes in determining the transfer efficiency.

\begin{figure*}[tbp]
	\center
	\includegraphics[width=1 \textwidth]{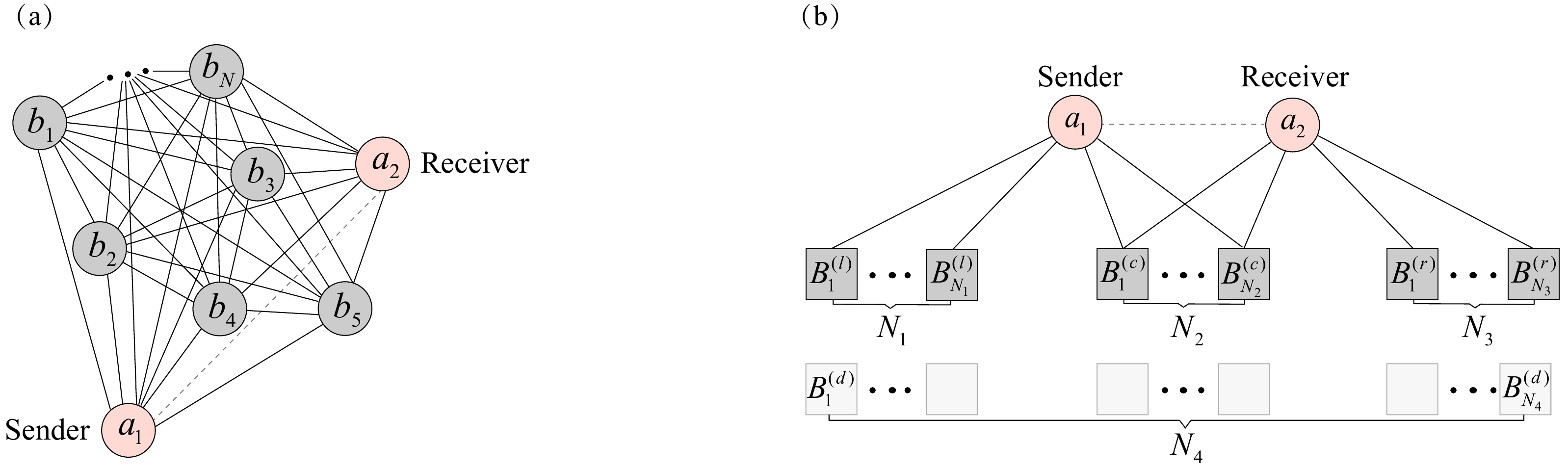}
	\caption{Configurations of the coupled bosonic network in both the original and bipartite-graph representations. (a) The original representation of the coupled bosonic network consisting of  the sender mode $a_{1}$, the receiver mode $a_{2}$, and $N$ intermediate bosonic modes \{$b_{1}$,$b_{2},\ldots,b_{N}$\}. (b) The bipartite-graph coupling configuration among the sender mode $a_{1}$, the receiver mode $a_{2}$, and these intermediate normal modes. Here,  $B_{1\text{-}N_{1}}^{(l)}$ and $B_{1\text{-}N_{3}}^{(r)}$ are the intermediate  normal modes only coupled to the sender mode $a_{1}$ and the receiver mode $a_{2}$, respectively. $B_{1\text{-}N_{2}}^{(c)}$ and $B_{1\text{-}N_{4}}^{(d)}$ are the intermediate normal modes commonly coupled with and decoupled from the two modes $a_{1}$ and $a_{2}$, respectively. In this network, the total number of the intermediate normal modes is $N=N_{\text{c}}+N_{4}$ with $N_{\text{c}}=N_{1}+N_{2}+N_{3}$ being the number of the normal modes connected to the sender-receiver subnetwork and $N_{4}$ the number of decoupled modes. Since these $N_{4}$ decoupled normal modes have no influence on the excitation transfer, they can be ignored during the analyses of the excitation transfer. As a result, the bipartite-graph network can be conveniently labeled as $(N_{1},N_{2},N_{3}|\text{Yes or No})$, where $N_{1}$, $N_{2}$, and $N_{3}$ represent the numbers of the intermediate normal modes coupled only to the sender mode $a_{1}$, only to the receiver mode $a_{2}$, and to both of them, respectively; and the Boolean variable ``Yes'' (``No'') indicates that the sender and the receiver are directly connected (disconnected).}
	\label{system_model}
\end{figure*}

In this paper, we propose a bipartite-graph framework to study quantum excitation transfer between two nodes in a ($2+N$)-mode coupled bosonic network. Here, the starting and ending nodes of the excitation transfer are denoted as the sender and receiver, respectively, and the remaining $N$ nodes form an intermediate subnetwork.  Further, we establish a bipartite-graph framework by diagonalizing the intermediate subnetwork into $N$ independent normal modes. 
The bipartite-graph framework can be understood as an application of the network deformation. It can be used to simplify the network configurations and provide a unified representation to regularize the network structure. In addition, it enables us to uncover the underlying physical mechanisms governing the excitation transfer in the original network, including the role of dark modes decoupled from both the sender and the receiver. Moreover, this framework enables the determination of the parameters for optimal transfer and offers the insights for inverse design of complex quantum networks. 

To investigate the quantum excitation transfer from the sender to the receiver in the coupled bosonic network, we derive the equation of motion for the covariance matrix and establish the relationship between the covariance matrices in the original and bipartite-graph representations. Concretely,
we analyze quantum excitation transfer in the bipartite-graph network for two cases: finite and infinite intermediate normal modes. For the case of finite intermediate normal modes, we consider the ($2+N$)-mode bipartite-graph network with $N=1,2,\ldots,5$, and we show how the transfer efficiency depends on the network configurations and system parameters. For the case of infinite intermediate normal modes, these normal modes can be considered as the effective vacuum baths containing discrete bosonic modes. 
Under the Weisskopf-Wigner approximation, we derive the quantum master equation and obtain the relationship between the transfer efficiency and the effective decay rates for various coupling configurations. By analyzing the quantum excitation transfer in the bipartite-graph network, we obtain the system parameters for the highly efficient quantum transfer of the excitations in various coupling configurations, and then the parameters corresponding to a highly efficient quantum transfer in the original bosonic network can be obtained accordingly. 

In this case, the bipartite-graph framework will not only provide a new insight for studying quantum resource transfer in complex and large-scale coupled quantum networks, but also establish a new paradigm for studying coupled quantum networks. Therefore, this work paves the way toward the realization of highly efficient quantum resource transfer in scalable coupled quantum networks, and the optimal implementation of quantum networks.

The rest of this paper is organized as follows. In Sec.~\ref{CBN}, we introduce the physical model for coupled bosonic networks and present the Hamiltonians. Concretely, we present both the original and bipartite-graph representations for the coupled bosonic networks. In Sec.~\ref{CMCB}, we derive quantum Langevin equations for the networks expressed in these two representations. We also introduce the covariance matrices and their equations of motion. In Secs.~\ref{QET} and~\ref{stifinite} , we investigate quantum excitation transfer in the finite- and infinite-mode bipartite-graph networks, respectively. Finally, we present some discussions on the experimental implementation of this scheme and provide a brief conclusion of this work in Sec.~\ref{DAD}.

\section{Coupled bosonic network and Hamiltonians~\label{CBN}}
We consider a coupled bosonic network consisting of ($2+N$) bosonic
modes (each mode is a node), as shown in Fig. \ref{system_model}(a). Here two bosonic modes, denoted as the sender and the receiver,
are coupled to other $N$ bosonic modes. For clearly understanding the role of these nodes in the network, we denote the sender and receiver nodes as modes $a_{1}$ and $a_{2}$, respectively, while other bosonic-mode nodes are described as modes $b_{j}$ for $j=1,2,\ldots,N$. Hereafter, we will interchangeably write ``$1,2,\ldots,N$'' as ``$1\text{-}N$'' for brevity. In this network, all the modes $b_{j}$ play the role of intermediate modes between the sender and receiver modes $a_{1}$ and $a_{2}$. We assume that all these bosonic modes could be coupled with each other via the excitation-exchange interactions. Concretely, the two modes $a_{1}$ and $a_{2}$ could be coupled with each other via a direct interaction, and all other intermediate modes $b_{j}$ can form a subnetwork. In addition, there exist couplings between the modes $a_{i=1,2}$ and $b_{j=1\text{-}N}$. 

Based on the above analyses, the Hamiltonian of the coupled bosonic networks can be written as 
\begin{equation}
		\hat{H}_{\text{cbn}}=	\hat{H}_{a}+	\hat{H}_{b}+	\hat{H}_{ab},\label{H000}
\end{equation}
where $	\hat{H}_{a}$, $	\hat{H}_{b}$, and $	\hat{H}_{ab}$ describe the Hamiltonians of the mode-$a$ subnetwork (including the sender and receiver modes $a_{1}$ and $a_{2}$), the mode-$b$ subnetwork (including these intermediate modes $b_{j=1\text{-}N}$), and the interactions between the nodes of the two subnetworks, respectively. Concretely, these Hamiltonians can be written as 
\begin{subequations}
	\begin{align}
			\hat{H}_{a} & =\omega_{\text{1}}\hat{a}_{1}^{\dagger}\hat{a}_{1}+\omega_{2}\hat{a}_{2}^{\dagger}\hat{a}_{2}+g \hat{a}_{1}^{\dagger}\hat{a}_{2}+g^{*}\hat{a}_{2}^{\dagger}\hat{a}_{1},\\
			\hat{H}_{b} & =\sum_{j=1}^{N}\omega_{b,j}\hat{b}_{j}^{\dagger}\hat{b}_{j}+\sum_{j,j^{\prime}=1,j<j^{\prime}}^{N}(\xi_{j,j^{\prime}}\hat{b}_{j}^{\dagger}\hat{b}_{j^{\prime}}+\xi_{j,j^{\prime}}^{*}\hat{b}_{j^{\prime}}^{\dagger}\hat{b}_{j}),\label{Hori}\\
			\hat{H}_{ab} & =\sum_{l=1}^{2}\sum_{j=1}^{N}(\eta_{l,j}\hat{a}_{l}^{\dagger}\hat{b}_{j}+\eta_{l,j}^{*}\hat{b}_{j}^{\dagger}\hat{a}_{l}),
	\end{align}
\end{subequations}
where $\hat{a}_{1}$ ($\hat{a}_{1}^{\dagger}$) and $\hat{a}_{2}$ ($\hat{a}_{2}^{\dagger}$)
are, respectively, the annihilation (creation) operators of the sender and receiver modes, with the corresponding resonance frequencies $\omega_{\text{1}}$ and
$\omega_{\text{2}}$, and $g$ is the coupling strength between
the sender and receiver modes. In addition, $\hat{b}_{j}$ ($\hat{b}_{j}^{\dagger}$) is the
annihilation (creation) operator of the $j$th bosonic
mode $b_{j}$ with the resonance frequency $\omega_{b,j}$, and $\xi_{jj^{\prime}}$
is the hopping coupling strength between the two bosonic modes $b_{j}$
and $b_{j^{\prime}}$. The variable $\eta_{1,j}$ ($\eta_{2,j}$) is the
coupling strength between the sender (receiver) mode $a_{1}$ ($a_{2}$) and the $j$th intermediate bosonic
mode $b_{j}$.

To be concise, we introduce the operator vectors for these intermediate modes as
\begin{eqnarray}
	\hat{\textbf{b}}=(\hat{b}_{1},\hat{b}_{2},\ldots,\hat{b}_{N})^{\text{T}},\hspace{0.5cm}\hat{\textbf{b}}^{\dagger}=(\hat{b}^{\dagger}_{1},\hat{b}^{\dagger}_{2},\ldots,\hat{b}^{\dagger}_{N}),
\end{eqnarray}
where ``\text{T}" denotes the matrix transpose.
Equation (\ref{H000}) can then be expressed as a compact form
\begin{equation}	
	\hat{H}_{\text{cbn}}=(\hat{a}^{\dagger}_{1},\hat{a}^{\dagger}_{2},\hat{\textbf{b}}^{\dagger})\textbf{H}^{(N)}_{ab}\left(\begin{array}{c}
		\hat{a}_{1}	\\
	\hat{a}_{2}	\\
		\hat{\textbf{b}}	\\
	\end{array}\right),\label{Habab}
\end{equation}
where we introduce the coefficient matrix
\begin{equation}
	\textbf{H}^{(N)}_{ab}=\left(\begin{array}{c|c}
		\textbf{H}_{a}&	\textbf{C}_{ab}\\\hline
		\textbf{C}^{\dagger}_{ab}&	\textbf{H}_{b}
	\end{array}\right)=\left(\begin{array}{cc|cccc}
		\omega_{\text{1}} & g & \eta_{1,1} & \eta_{1,2} & \dotsb & \eta_{1,N}\\
		g^{*} & \omega_{\text{2}} & \eta_{2,1} & \eta_{2,2} & \dotsb & \eta_{2,N}\\
		\hline \eta_{1,1}^{*} & \eta_{2,1}^{*} & \omega_{b,1} & \xi_{1,2} &  \dotsb & \xi_{1,N}\\
		\eta_{1,2}^{*} & \eta_{2,2}^{*} & \xi_{1,2}^{*} & \omega_{b,2} &  \dotsb & \xi_{2,N}\\
		\vdots & \vdots & \vdots & \vdots & \ddots & \vdots\\
		\eta_{1,N}^{*} & \eta_{2,N}^{*} & \xi_{1,N}^{*} & \xi_{2,N}^{*} &  \dotsb & \omega_{b,N}
	\end{array}\right).~\label{Habab}
\end{equation}
Note that we add the superscript “$(N)$” in $\textbf{H}^{(N)}_{ab}$ to denote the number of the intermediate modes.

From the viewpoint of quantum information science, the implementation of a highly efficient quantum resource transfer from the sender to the receiver in the network described in Fig.~\ref{system_model}(a) is a significant task. In particular, how to find the optimal parameter conditions, optimize the transmission paths, and achieve a high transmission efficiency are crucial for implementing highly efficient quantum resource transfer in this network.  

We can see from Eq. (\ref{Habab}) that the maximal distance between the sender and the receiver is $N+1$ for the coupled bosonic network depicted in Fig.~\ref{system_model}(a).  The large distance makes it difficult to analyze the optimal transfer condition. To address this issue, we propose a bipartite-graph framework to study quantum excitation transfer in the networks. Concretely, we diagonalize the intermediate-mode network by introducing a unitary operator $\textbf{U}$, i.e., $\textbf{H}_{B}=\textbf{U}\textbf{H}_{b}\textbf{U}^{\dagger}	=\text{diag}(\Omega_{1},\Omega_{2},\ldots,\Omega_{N})$,
where $\Omega_{j}$ is the resonance frequency of the $j$th intermediate normal mode defined by
\begin{equation}
	B_{j}=\sum_{j^{\prime}=1}^{N}\textbf{U}_{j,j^{\prime}}b_{j^{\prime}},\hspace{0.5cm}j=1,2,\ldots,N.~\label{BjU}
\end{equation}
Here, the symbol $``\text{diag}(\cdot)"$ represents the generation of a diagonal matrix with the elements of $(\cdot)$ on the main diagonal. Using the operator $\textbf{U}$, the intermediate-mode subnetwork can be transformed into $N$ independent intermediate normal modes $B_{j=1\text{-}N}$.
In this case,  the coupled bosonic network depicted in Fig. \ref{system_model}(a) can be transformed into the bipartite-graph network, as shown in Fig. \ref{system_model}(b).  For clarity, we show this transformation using a (2+2)-mode network that contains two degenerate type-$b$ modes $\{b_{1},b_{2}\}$ coupled via a real hopping amplitude $\xi_{1,2}$. The corresponding coefficient matrix is given in Eq.~(\ref{Habab}). We can diagonalize $\textbf{H}_{b}$ by introducing the unitary matrix $\mathbf{U}=\frac{1}{\sqrt{2}}\begin{pmatrix} -1 & 1 \\ 1 & 1 \end{pmatrix}$, and then the intermediate-mode subnetwork can be transformed into two independent normal modes $B_{j}=[b_{2}+(-1)^{j}b_{1}]/\sqrt{2}$ with the normal-mode frequencies $\Omega_{j}=\omega_{m}+(-1)^{j}\xi_{1,2}$ for $j=1,2$.  This diagonalization process effectively reshapes the network configuration: the original complex connections are simplified into a bipartite-graph configuration where the sender and receiver modes interact exclusively with two independent normal modes $\{B_{1},B_{2}\}$, thereby providing a clear visualization of network deformation.

For the bipartite-graph network, there are two connected subnetworks $\{a_{1},a_{2}\}$ and $\{B_{j=1\text{-}N}\}$. The former subnetwork consists of the sender mode $a_{1}$ and the receiver mode $a_{2}$, while the latter consists of $N$ intermediate normal modes $B_{j=1\text{-}N}$. The  intermediate normal-mode subnetwork can be divided into four subsets: the first subset contains $N_{1}$ normal modes \{$B_{1}^{(l)},\ldots,B_{N_{1}}^{(l)}$\} connected only to the sender mode $a_{1}$; the second subset includes $N_{2}$ normal modes \{$B_{1}^{(c)},\ldots,B_{N_{2}}^{(c)}$\} coupled to both the sender mode $a_{1}$  and the receiver mode $a_{2}$; the third subset consists of $N_{3}$ normal modes \{$B_{1}^{(r)},\ldots,B_{N_{3}}^{(r)}$\}  connected only to the receiver mode $a_{2}$; and the fourth subset has $ N_{4}$ normal modes \{$B_{1}^{(d)},\ldots,B_{N_{4}}^{(d)}$\}, which are decoupled from both the sender mode  $a_{1}$ and the receiver mode $a_{2}$. Therefore, there exists the relation $N=N_{\text{c}}+N_{4}$ with $N_{\text{c}}=N_{1}+N_{2}+N_{3}$ ($N_{4}$) being the total number of the normal modes coupled to (decoupled from) the sender-receiver subnetwork. Since these $N_{4}$ decoupled modes are not involved in the dynamical evolution of the network, they can be ignored in the following discussions. To be concise, we label the configuration in Fig.~\ref{system_model}(b) as $(N_{1},N_{2},N_{3}|\text{Yes or No})$, where the Boolean variable ``Yes'' (``No'') means the presence (absence) of a direct coupling $g$ between the sender and receiver modes. For example, the notation $(0,1,0|\text{Yes or No})$ denotes a single type-$b$ normal mode interacting with
both the sender mode $a_{1}$ and the receiver mode $a_{2}$. Note that, when $N_{1}=N_{3}=N_{4}=0$, the bipartite-graph network is reduced to a complete bipartite-graph network~\cite{west2001}.

According to the network structure, we can obtain the numbers associated with the coupling configurations. Concretely, when $N_{2}=0$, the quantum excitation transfer requires a direct coupling between the modes $a_{1}$ and $a_{2}$, then the number of the configurations is given by
\begin{equation}
		T_{1}=\binom{N_{\text{c}}+1}{1}=N_{\text{c}}+1,
	\end{equation}
	with the binomial coefficient $\binom{n}{m}=n!/[m!(n-m)!]$.
	In contrast, if $N_{2}>0$, the number of configurations is
\begin{equation}
		T_{2}=\binom{N_{\text{c}}+2}{2}-\binom{N_{\text{c}}+1}{1}=\frac{1}{2}N_{\text{c}}(N_{\text{c}}+1),
\end{equation}
	and this number needs to be doubled to cover the two cases corresponding to the presence and absence of the direct coupling between the two modes $a_{1}$ and $a_{2}$. By combining the above cases, the total number of all possible configurations can be obtained as 
	\begin{equation}
		N_{\text{total}}=T_{1}+2T_{2}=(N_{\text{c}}+1)^{2}
	\end{equation}
	for the bipartite-graph network.

We can also see from Fig.~\ref{system_model}(b) that the number of the couplings should be smaller or equal to $2N_{c}+1$ and the maximal distance between the sender and the receiver is reduced to 2. Owing to this feature, the bipartite-graph network can provide a clearer picture to explain the physical mechanism underlying the excitation transfer. When there is a direct connection between the sender and  the receiver, i.e., $g\neq0$, the bipartite-graph network becomes a semi-bipartite-graph network~\cite{xu2010}.  
For convenience, we sort these $N$ intermediate normal modes in Fig.~\ref{system_model}(b) so that $\textbf{H}_{B}$ can be expressed as 
\begin{eqnarray}
	\textbf{H}_{B}&=&\text{diag}(\Omega^{(l)}_{1},\ldots,\Omega^{(l)}_{N_{1}},\Omega^{(c)}_{1},\ldots,\Omega^{(c)}_{N_{2}},\Omega^{(r)}_{1},\ldots,\Omega^{(r)}_{N_{3}},\notag\\
	&&\Omega^{(d)}_{1},\ldots,\Omega^{(d)}_{N_{4}}),~\label{HB}
\end{eqnarray} 
where $\Omega_{s}^{(\cdot)}$ is the resonance frequency of the intermediate normal modes  corresponding to these four subsets.

To conveniently describe the coupled bosonic network depicted in Fig.~\ref{system_model}(b), we introduce the normal-mode operator vectors
\begin{subequations}~\label{BBdag}
	\begin{align}
		\hat{\textbf{B}}&\!=\!(\hat{B}^{(l)}_{1},\!\ldots\!,\hat{B}^{(l)}_{N_{1}},\hat{B}^{(c)}_{1},\!\ldots\!,\hat{B}^{(c)}_{N_{2}},\hat{B}^{(r)}_{1},\!\ldots\!,\hat{B}^{(r)}_{N_{3}},\hat{B}^{(d)}_{1},\!\ldots\!,\hat{B}^{(d)}_{N_{4}})^{\text{T}}\notag\\
		\quad&\!=\!\textbf{U}\hat{\textbf{b}},\\
		\hat{\textbf{B}}^{\dagger}&\!=\!(\hat{B}^{(l)\dagger}_{1},\!\ldots\!,\hat{B}^{(l)\dagger}_{N_{1}},\hat{B}^{(c)\dagger}_{1},\!\ldots\!,\hat{B}^{(c)\dagger}_{N_{2}},\hat{B}^{(r)\dagger}_{1},\!\ldots\!,\hat{B}^{(r)\dagger}_{N_{3}},\hat{B}^{(d)\dagger}_{1},\!\ldots\!,\hat{B}^{(d)\dagger}_{N_{4}})\notag\\
		\quad&\!=\!\textbf{b}^{\dagger}\textbf{U}^{\dagger}.
	\end{align}
\end{subequations} 
The Hamiltonian of the coupled bosonic network in the bipartite-graph framework can then be expressed as
\begin{equation}	
		\hat{H}_{\text{cbn}}=(\hat{a}^{\dagger}_{1},\hat{a}^{\dagger}_{2},\hat{\textbf{B}}^{\dagger})\textbf{H}^{(N)}_{aB}\left(\begin{array}{c}
		\hat{a}_{1}	\\
		\hat{a}_{2}	\\
		\hat{\textbf{B}}	\\
	\end{array}\right),\label{HaBaB}
\end{equation}
where we introduce the coefficient matrix in the bipartite-graph representation as
\begin{equation}
	\textbf{H}_{aB}^{(N)}=\left(\begin{array}{c|c}
		\textbf{H}_{a} & \textbf{C}_{aB}\\\hline
		\textbf{C}_{aB}^{\dagger} & \textbf{H}_{B}
	\end{array}\right).~\label{HaBN}
\end{equation}
In Eq.~(\ref{HaBN}), the submatrix $\textbf{H}_{a}$ is defined by 
$\textbf{H}_{a}=\left(\begin{array}{cc}
	\omega_{1} &g\\
		g^{*} &\omega_{2}
	\end{array}\right)$, $\textbf{H}_{B}$ is defined in Eq.~(\ref{HB}), and the coupling matrix $\textbf{C}_{aB}$ is defined by the elements 
\begin{equation}
	(\textbf{C}_{aB})_{k,j}=G_{k,j}=\sum_{j^{\prime}=1}^{N}\eta_{k,j^{\prime}}(\textbf{U}^{\dagger})_{j^{\prime},j}.~\label{CaBCaB}
\end{equation}
Note that, the effective coupling strength $G_{k,j}$ is a linear combination of the original couplings $\eta_{k,j^{\prime}}$, weighted by the elements of the unitary matrix $\mathbf{U}$. For the (2+2)-mode example above, the effective couplings are given by $G_{k,j}=[\eta_{k,2}+(-1)^{j}\eta_{k,1}]/\sqrt{2}$ for $j=1,2$.

According to the sorting order corresponding to Eq.~(\ref{HB}) and the coupling configuration in Fig.~\ref{system_model}(b), we know that the coupling matrix can be expressed as 
\begin{equation}
	\textbf{C}_{aB}=\left(\begin{array}{cccc}
		\textbf{G}_{1}^{(l)} &\textbf{G}_{1}^{(c)}&\textbf{0}_{1\times N_{3}}&\textbf{0}_{1\times N_{4}}\\
	\textbf{0}_{1\times N_{1}} &\textbf{G}_{2}^{(c)}&\textbf{G}_{2}^{(r)}&\textbf{0}_{1\times N_{4}}
	\end{array}\right).
\end{equation} 
Here, these submatrices are defined by
\begin{subequations}
	\begin{align}
 \textbf{G}_{1}^{(l)}&=(G_{1,1},G_{1,2},\ldots,G_{1,N_{1}}),\\ \textbf{G}_{1}^{(c)}&=(G_{1,N_{1}+1},G_{1,N_{1}+2},\ldots,G_{1,N_{1}+N_{2}}),\\ \textbf{G}_{2}^{(c)}&=(G_{2,N_{1}+1},G_{2,N_{1}+2},\ldots,G_{2,N_{1}+N_{2}}),\\ \textbf{G}_{2}^{(r)}&=(G_{2,N_{1}+N_{2}+1},G_{2,N_{1}+N_{2}+2},\ldots,G_{2,N_{1}+N_{2}+N_{3}}),
	\end{align}
\end{subequations}
\\and $\textbf{0}_{1\times N_{s}}=(0,0,\ldots,0)_{1\times N_{s}}$ for $s=1,3,4$. These matrix elements have been introduced in Eq.~(\ref{CaBCaB}).

\section{Covariance matrix of the coupled bosonic networks~\label{CMCB}}
For a practical network, its environment will inevitably affect the efficiency of quantum excitation transfer. To include the dissipation effect, we adopt the quantum Langevin equations (QLEs) to describe the evolution of the nodal modes in the networks. To analyze the quantum excitation transfer, we need to derive the equation of motion for the covariance matrix (CM). In this section, we will present a detailed derivation of the equation of motion for the CM in both the original and bipartite-graph representations of the coupled bosonic network. 

\subsection{Covariance matrix of the coupled bosonic networks in the original representation}
To include the dissipation of the coupled bosonic network depicted in Fig.~\ref{system_model}(a), we assume that all the nodal modes in the  network are coupled to individual vacuum baths. By introducing the operator vectors 
\begin{subequations}
	\begin{align}
		\hat{\textbf{u}}(t) &\! =\!  [\hat{a}_{1}(t),\hat{a}_{2}(t),\hat{b}_{1}(t),\!\ldots\!,\hat{b}_{N}(t),\hat{a}_{1}^{\dagger}(t),\hat{a}_{2}^{\dagger}(t), \hat{b}_{1}^{\dagger}(t),\!\ldots\!,\hat{b}_{N}^{\dagger}(t)]^{T}, \\
		\hat{\textbf{N}}(t) & \!=\!  [\hat{F}_{a_{1}}(t),\hat{F}_{a_{2}}(t),\hat{F}_{b_{1}}(t),\!\ldots\!,\hat{F}_{b_{N}}(t),\hat{F}_{a_{1}}^{\dagger}(t),\hat{F}_{a_{2}}^{\dagger}(t),\notag\\
 &\quad F_{b_{1}}^{\dagger}(t),\!\ldots\!,F_{b_{N}}^{\dagger}(t)]^{T},~\label{Fa1b}
	\end{align}~\label{uNt}
\end{subequations}
\\the QLEs of the nodal modes in the network can be written as
\begin{equation}
	\dot{\hat{\textbf{u}}}(t)=\textbf{A}\hat{\textbf{u}}(t)+\hat{\textbf{N}}(t),\label{udot}
\end{equation}
where we introduce the drift  matrix $\textbf{\ensuremath{\textbf{A}}}=\left(\begin{array}{cc}
	-\textbf{E} & \textbf{0}\\
	\textbf{0} & -\textbf{E}^{*}
\end{array}\right)$,
with
\small{
\begin{align}
	\textbf{E} & =\left(\begin{array}{cccccc}
		i\omega_{\text{1}}+\frac{\kappa_{1}}{2} & ig & i\eta_{1,1} & i\eta_{1,2} & \ldots & i\eta_{1,N}\\
		ig^{*} & i\omega_{2}+\frac{\kappa_{2}}{2} & i\eta_{2,1} & i\eta_{2,2} & \ldots & i\eta_{2,N}\\
		i\eta_{1,1}^{*} & i\eta_{2,1}^{*} & i\omega_{b,1}+\frac{\gamma_{1}}{2} & i\xi_{1,2} & \ldots & i\xi_{1,N}\\
		i\eta_{1,2}^{*} & i\eta_{2,2}^{*} & i\xi_{1,2}^{*} & i\omega_{b,2}+\frac{\gamma_{2}}{2} & \ldots & i\xi_{2,N}\\
		\vdots & \vdots & \vdots & \vdots & \ddots & \vdots\\
		i\eta_{1,N}^{*} & i\eta_{2,N}^{*} & i\xi_{1,N}^{*} & i\xi_{2,N}^{*} & \ldots & i\omega_{b,N}+\frac{\gamma_{N}}{2}
	\end{array}\right).~\label{textb}
\end{align}}In Eq.~(\ref{Fa1b}), $\hat{F}_{a_{1}}$, $\hat{F}_{a_{2}}$, and $\hat{F}_{b_{j=1\text{-}N}}$ are, respectively, the noise operators of the baths associated with the sender, the receiver, and the intermediate modes $b_{j=\text{1-}N}$. The variables in Eq.~(\ref{textb}) have been defined in Eq.~(\ref{H000}). These noise operators satisfy these nonzero correlation functions
\begin{subequations}~\label{correlationb}
	\begin{align}
		\langle \hat{F}_{a_{l}}(t)\hat{F}_{a_{l}}^{\dagger}(t^{\prime})\rangle  &=\kappa_{l}(\bar{n}_{l}+1)\delta(t-t^{\prime}),\\
		\langle \hat{F}_{a_{l}}^{\dagger}(t)\hat{F}_{a_{l}}(t^{\prime})\rangle   &=\kappa_{l}\bar{n}_{l}\delta(t-t^{\prime}), \\
		\langle \hat{F}_{b_{j}}(t)\hat{F}_{b_{j}}^{\dagger}(t^{\prime})\rangle   &=\gamma_{j}(\bar{n}_{b,j}+1)\delta(t-t^{\prime}),\\
		\langle \hat{F}_{b_{j}}^{\dagger}(t)\hat{F}_{b_{j}}(t^{\prime})\rangle  
		&=\gamma_{j}\bar{n}_{b,j}\delta(t-t^{\prime}),
	\end{align}
\end{subequations}
for $l=1,2$ and $j=1\text{-}N$, where $\bar{n}_{l}$ and $\bar{n}_{b,j}$ denote the average thermal excitation numbers associated with the baths of these modes $a_{l}$ and $b_{j}$, respectively. The variables $\kappa_{1}$, $\kappa_{2}$, and $\gamma_{j=1\text{-}N}$ are
the decay rates corresponding to the sender, the receiver, and the modes $b_{j=\text{1-}N}$.

To study quantum transfer of a single excitation from the sender to the receiver in the bosonic network, we need to calculate the mean excitation number of the receiver mode $a_{2}$. For our purpose, we introduce the covariance matrix defined by the elements
\begin{equation}
\textbf{R}_{l,l^{\prime}}(t)=\frac{1}{2}\left[\left\langle \hat{\textbf{\ensuremath{\textbf{u}}}}_{l}(t)\hat{\textbf{\ensuremath{\textbf{u}}}}_{l^{\prime}}(t)\right\rangle +\left\langle \hat{\textbf{u}}_{l^{\prime}}(t)\hat{\textbf{\ensuremath{\textbf{u}}}}_{l}(t)\right\rangle \right]~\label{RLLt}
\end{equation}
for $l,l^{\prime}=1,2,\ldots,(N+2)$.
The evolution of the covariance matrix is governed by the equation of motion
\begin{eqnarray}
	\dot{\textbf{R}}(t) & = & \textbf{\ensuremath{\textbf{A}}}\textbf{R}(t)+\textbf{R}(t)\textbf{A}^{T}+\textbf{C},\label{Rtdottt}
\end{eqnarray}
where we introduce $\textbf{C}=\left(\begin{array}{cc}
	\textbf{0} &\textbf{P}\\
	\textbf{P} & \textbf{0}
\end{array}\right)$
with
\small{
\begin{equation}
		\textbf{\ensuremath{\textbf{P}}}=\left(\begin{array}{ccccc}
			\kappa_{1}(\bar{n}_{1}+\frac{1}{2}) & 0 & 0 & \dotsb & 0\\
			0 & \kappa_{2}(\bar{n}_{2}+\frac{1}{2}) & 0 &  \dotsb & 0\\
			0 & 0 & \gamma_{1}(\bar{n}_{b,1}+\frac{1}{2}) & \dotsb & 0\\
			\vdots & \vdots & \vdots & \ddots & \vdots\\
			0 & 0 & 0 & \dotsb & \gamma_{N}(\bar{n}_{b,N}+\frac{1}{2})
		\end{array}\right).~\label{PP}
\end{equation} }By solving Eq. (\ref{Rtdottt}), we can obtain the expression of the covariance matrix as
\begin{equation}
	\textbf{R}(t)=\textbf{G}(t)\textbf{R}(0)\textbf{G}^{T}(t)+\textbf{G}(t)\textbf{Z}(t)\textbf{G}^{T}(t),~\label{RGtt}
\end{equation}
where
\begin{equation}
	\textbf{Z}(t)=\int_{0}^{t}\textbf{G}^{-1}(\tau)\textbf{C}\left[\textbf{G}^{-1}(\tau)\right]^{T}d\tau,
\end{equation}
with $\textbf{G}(t)=\exp(\textbf{A}t)\textbf{G}(0)$ and $\textbf{G}(0)=\textbf{1}$.
Then the mean excitation number of the receiver mode $a_{2}$ at time $t$ can be expressed as
\begin{equation}
	\langle \hat{a}_{2}^{\dagger}\hat{a}_{2}(t)\rangle =\textbf{R}_{N+4,2}(t)-\frac{1}{2},~\label{a2daga2}
\end{equation}
where $\textbf{R}_{N+4,2}(t)$ is the matrix element defined in Eq. (\ref{RLLt}).

Following the above-mentioned procedure, we can obtain the result of the quantum excitation transfer. However, to identify the influence of quantum interference effect on quantum excitation transfer, we need to analyze the equation of motion for the covariance matrix in the bipartite-graph representation. To this end, below we perform a detailed calculation of the excitation transfer in the bipartite-graph representation.

\subsection{Covariance matrix of the coupled bosonic networks in the bipartite-graph representation}
For the bipartite-graph network depicted in Fig.~\ref{system_model}(b), the intermediate normal mode $B_{j}$ is composed of these intermediate modes $b_{j=1\text{-}N}$, and the relation between the operator vectors $\textbf{B}$ and $\textbf{b}$ is given in Eq.~(\ref{BBdag}). For the normal mode $B_{j}$, it is not only affected by the bath associated with $b_{j}$, but will also be affected by these baths coupled with other intermediate modes $b_{j^{\prime}\neq j}$. Accordingly, we introduce the noise operator $\hat{F}_{B_{j}}(t)$ for the normal mode $B_{j}$ as
\begin{equation}
	\hat{F}_{B_{j}}(t)=\sum_{j^{\prime}=1}^{N}\textbf{U}_{j,j^{\prime}}\hat{F}_{b_{j^{\prime}}}(t).\label{binj}
\end{equation}
Based on Eqs.~(\ref{correlationb}) and (\ref{binj}), we can obtain the correlation functions for these normal-mode noise operators $\hat{F}_{B_{j}}(t)$ as
\begin{subequations}~\label{Bincorre}
	\begin{align}
		\langle \hat{F}_{B_{j}}(t)\hat{F}_{B_{j}}^{\dagger}(t^{\prime})\rangle  & =\sum_{j^{\prime}=1}^{N}\textbf{U}_{j,j^{\prime}}	\langle\hat{F}_{b_{j^{\prime}}}(t)\hat{F}_{b_{j^{\prime}}}^{\dagger}(t^{\prime})\rangle(\textbf{U}^{\dagger})_{j^{\prime},j},\\
		\langle \hat{F}_{B_{j}}^{\dagger}(t)\hat{F}_{B_{j}}(t^{\prime})\rangle  & =\sum_{j^{\prime}=1}^{N}\textbf{U}_{j,j^{\prime}}	\langle \hat{F}_{b_{j^{\prime}}}^{\dagger}(t)\hat{F}_{b_{j^{\prime}}}(t^{\prime})\rangle(\textbf{U}^{\dagger})_{j^{\prime},j} .
	\end{align}
\end{subequations}
Note that the correlation functions in Eqs.~(\ref{Bincorre}) are the superposition of nonzero correlation functions of the noise operators $\hat{F}_{b_{j}}(t)$, where the superposition coefficients are determined by the unitary matrix $\textbf{U}$. 

To analyze the quantum excitation transfer in the bipartite-graph network, we further introduce the operator vectors 
\begin{subequations}
	\begin{align}
		\hat{\tilde{\textbf{u}}}(t) & =  [\hat{a}_{1}(t),\hat{a}_{2}(t),\hat{B}_{1}(t),\ldots,\hat{B}_{N}(t),\hat{a}_{1}^{\dagger}(t),\hat{a}_{2}^{\dagger}(t), \hat{B}_{1}^{\dagger}(t),\ldots,\hat{B}_{N}^{\dagger}(t)]^{T}\notag\\
		\quad &=\textbf{V}\hat{\textbf{u}}(t),\\
		\hat{\tilde{\textbf{N}}}(t) & \!=\! [\hat{F}_{a_{1}}(t),\hat{F}_{a_{2}}(t),\hat{F}_{B_{1}}(t),\!\ldots\!,\hat{F}_{B_{N}}(t),\hat{F}_{a_{1}}^{\dagger}(t),\hat{F}_{a_{2}}^{\dagger}(t),\hat{F}_{B_{1}}^{\dagger}(t),\!\ldots\!,\hat{F}_{B_{N}}^{\dagger}(t)]^{T}\notag\\	
		\quad &=\textbf{V}\hat{\textbf{N}}(t),~\label{tilNt}
	\end{align}
\end{subequations}
where the matrix $\textbf{V}$ can be expressed as 
\begin{equation}
	\textbf{V}=\textbf{I}_{2}\oplus \textbf{U}\oplus \textbf{I}_{2}\oplus \textbf{U}^{*},~\label{textbV}
\end{equation}
with the notation ``$\oplus$'' denoting the direct sum of matrices.
In Eq.~(\ref{textbV}), the matrix $\textbf{I}_{2}$ represents the $2\times2$ identity matrix and $\textbf{U}$ is given in Eq.~(\ref{BjU}). Then the QLEs can be written as
\begin{equation}
	\dot{\hat{\tilde{\textbf{u}}}}(t)=\tilde{\textbf{A}}\hat{\tilde{\textbf{u}}}(t)+\hat{\tilde{\textbf{N}}}(t),\label{udot-1}
\end{equation}
where we introduce the drift  matrix $\textbf{\ensuremath{\tilde{\textbf{A}}}}=\left(\begin{array}{cc}
	-\tilde{\textbf{E}} & \textbf{0}\\
	\textbf{0} & -\tilde{\textbf{E}}^{*}
\end{array}\right)$ with
\small{
\begin{equation}
\tilde{\textbf{\text{E}}}=\left(\begin{array}{cccccc}
	i\omega_{\text{1}}+\frac{\kappa_{1}}{2} & ig & iG_{1,1} & iG_{1,2} &  \dotsb & iG_{1,N}\\
	ig^{*} & i\omega_{2}+\frac{\kappa_{2}}{2} & iG_{2,1} & iG_{2,2} &  \dotsb& iG_{2,N}\\
	iG_{1,1}^{*} & iG_{2,1}^{*} & i\Omega_{1}+\frac{\tilde{\boldsymbol{\Gamma}}{}_{1,1}}{2} & \frac{\tilde{\boldsymbol{\Gamma}}{}_{1,2}}{2} & \dotsb& \frac{\tilde{\boldsymbol{\Gamma}}{}_{1,N}}{2}\\
	iG_{1,2}^{*} & iG_{2,2}^{*} & \frac{\tilde{\boldsymbol{\Gamma}}{}_{2,1}}{2} & i\Omega_{2}+\frac{\tilde{\boldsymbol{\Gamma}}{}_{2,2}}{2} & \dotsb & \frac{\tilde{\boldsymbol{\Gamma}}{}_{2,N}}{2}\\
	\vdots & \vdots & \vdots & \vdots & \ddots & \vdots\\
	iG_{1,N}^{*} & iG_{2,N}^{*} & \frac{\tilde{\boldsymbol{\Gamma}}{}_{N,1}}{2} & \frac{\tilde{\boldsymbol{\Gamma}}{}_{N,2}}{2} & \dotsb & i\Omega_{N}+\frac{\tilde{\boldsymbol{\Gamma}}{}_{N,N}}{2}
\end{array}\right).~\label{tilE}
\end{equation}}

\noindent In Eq.~(\ref{tilE}), $\tilde{\boldsymbol{\Gamma}}_{j,j^{\prime}}$ are the elements of $\tilde{\boldsymbol{\Gamma}}=\textbf{U}\boldsymbol{\Gamma}\textbf{U}^{\dagger}$ with the damping matrix $\boldsymbol{\Gamma}=\text{diag}(\gamma_{1},\gamma_{2},\ldots,\gamma_{N})$.

Similarly, we can analyze the quantum excitation transfer by calculating the mean excitation numbers based on Eq. (\ref{udot-1}). To this end, we introduce the covariance matrix defined by the elements
\begin{equation}
	\tilde{\textbf{R}}_{l,l^{\prime}}(t)=\frac{1}{2}[\langle \hat{\tilde{\textbf{u}}}_{l}(t)\hat{\tilde{\textbf{u}}}_{l^{\prime}}(t)\rangle +\langle \hat{\tilde{\textbf{u}}}_{l^{\prime}}(t)\hat{\tilde{\textbf{u}}}_{l}(t)\rangle ],\label{RLLp}
\end{equation}
which satisfies the equation of motion
\begin{eqnarray}
	\dot{\tilde{\textbf{R}}}(t) & = & \textbf{\ensuremath{\tilde{\textbf{A}}}}\tilde{\textbf{R}}(t)+\tilde{\textbf{R}}(t)\tilde{\textbf{A}}^{T}+\tilde{\textbf{C}}.\label{CMEE}
\end{eqnarray}
In Eq.~(\ref{CMEE}), we introduce the matrix $\tilde{\textbf{C}}=\left(\begin{array}{cc}
	\textbf{0} & \tilde{\textbf{P}}\\
	\tilde{\textbf{P}} & \textbf{0}
\end{array}\right)$,
where the submatrix $\tilde{\textbf{P}}$ takes the form
\begin{equation}
\textbf{\ensuremath{\tilde{\textbf{P}}}}=(\textbf{I}_{2}\oplus\textbf{U})\textbf{P}(\textbf{I}_{2}\oplus\textbf{U}^\dagger)~\label{tildaP}
\end{equation} 	
with $\textbf{P}$ given in Eq.~(\ref{PP}). When the original damping rates are nonidentical (i.e., $\gamma_{j}\neq\gamma_{j^{\prime}}$), $\tilde{\textbf{P}}$ is generally nondiagonal, and Eq.~(\ref{tildaP}) retains the cross-noise correlations between different normal modes.
Then the mean excitation number of the receiver mode $a_{2}$  is given by
\begin{equation}
	\langle \hat{a}_{2}^{\dagger}\hat{a}_{2}(t)\rangle =\tilde{\textbf{R}}_{N+4,2}(t)-\frac{1}{2},~\label{a2daga2}
\end{equation}
where $\tilde{\textbf{R}}_{N+4,2}(t)$ can be obtained by solving Eq.~(\ref{CMEE}). Based on Eq.~(\ref{BjU}), we can obtain the relation between the covariance matrices $\tilde{\textbf{R}}(t)$ and $\textbf{R}(t)$ as
\begin{equation}
	\tilde{\textbf{R}}(t)=\textbf{V}\textbf{R}(t)\textbf{V}^{T},
\end{equation}
where $\textbf{V}$ is given in Eq.~(\ref{textbV}).

In the following two sections, we will analyze quantum excitation transfer from the sender mode $a_{1}$ to the receiver mode $a_{2}$ within the bipartite-graph framework. 

\begin{figure}[tbp]
	\center
	\includegraphics[width=0.48 \textwidth]{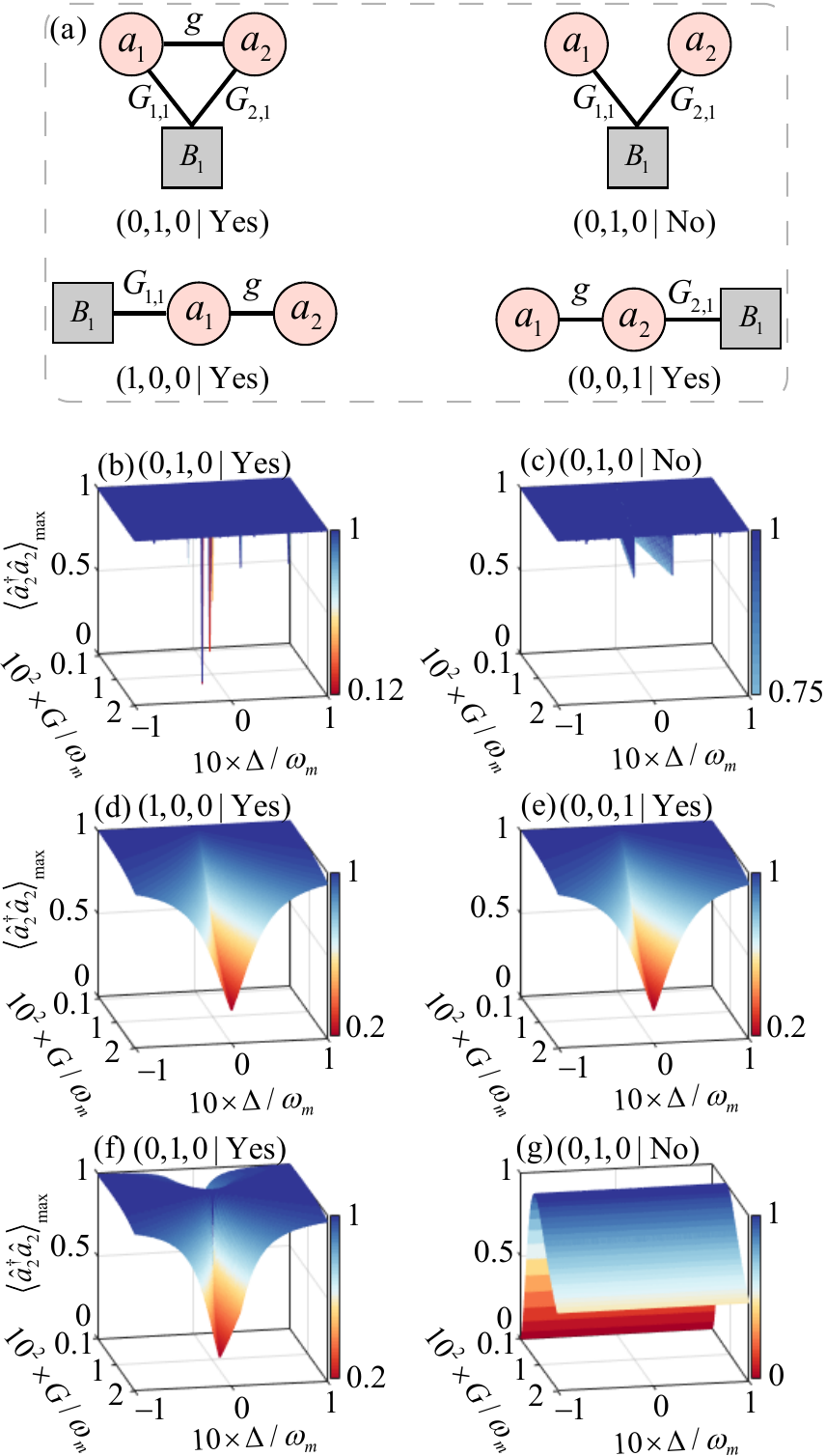}
	\caption{(a) Four coupling configurations of the ($2+1$)-mode bipartite-graph network. [(b)-(g)] Maximum mean excitation number $\langle \hat{a}^{\dagger}_{2}\hat{a}_{2}\rangle_{\text{max}}$ vs the coupling strengths $G_{1,1}/\omega_{m}, G_{2,1}/\omega_{m}$, and the detuning $\Delta/\omega_{m}$ in these cases: (b) $g=0.01\omega_{m}, G_{1,1}=G_{2,1}=G$; (c) $g=0, G_{1,1}=G_{2,1}=G$; (d) $g=0.01\omega_{m}, G_{2,1}=0, G_{1,1}=G$; (e) $g=0.01\omega_{m}, G_{1,1}=0, G_{2,1}=G$; (f) $g=0.01\omega_{m}, G_{1,1}=0.008\omega_{m}, G_{2,1}=G$; and (g) $g=0, G_{1,1}=0.008\omega_{m}, G_{2,1}=G$ in the ($2+1$)-mode bosonic networks. The notation $(N_{1},N_{2},N_{3}|\text{Yes or No})$ is introduced in Fig.~\ref{system_model}. Other common parameters used are $\omega_{1}=\omega_{2}=\omega_{m}$.}
	\label{mode_2_1}
\end{figure}
\section{Quantum excitation transfer in the ($2+N_{\text{c}}$)-mode bipartite-graph network with finite values of $N_{\text{c}}$~\label{QET}}
In coupled bosonic networks, quantum excitation transfer provides a fundamental mechanism for transmitting physical carriers, such as photons or phonons, between any two nodes. In this work, we quantify the transfer performance by the average excitation number received by the mode $a_{2}$, focusing on excitation transfer rather than full quantum-state transfer fidelity.
Specifically, we analyze quantum excitation transfer dynamics of a coupled bosonic network in the bipartite-graph framework by employing the covariance-matrix formalism. Since the total excitation number operator $\hat{a}_{1}^{\dagger}\hat{a}_{1}+\hat{a}_{2}^{\dagger}\hat{a}_{2}+\sum_{j=1}^{N}\hat{B}_{j}^{\dagger}\hat{B}_{j}$ is a conserved quantity, we can restrict the system within the single-excitation subspace when we consider that the single excitation is initially in the sender mode $a_{1}$. In this case, the initial covariance matrix elements in Eq.~(\ref{RLLp})  are given by $\tilde{R}_{N_{\text{c}}+3,1}(0)=\tilde{R}_{1,N_{\text{c}}+3}(0)=3/2$ and $\tilde{R}_{N_{\text{c}}+2+j,j}(0)=\tilde{R}_{j,N_{\text{c}}+2+j}(0)=1/2$ for $j=2,3,\ldots,N_{\text{c}}+2$, and all other elements are zero. Next, we will analyze the transfer of the single excitation from the sender mode $a_{1}$ to the receiver mode $a_{2}$ in the presence of different $N_{\text{c}}$ intermediate modes.

\subsection{Quantum excitation transfer in the ($2+1$)-mode bipartite-graph network}
In the case of $N_{\text{c}}=1$, i.e., for the ($2+1$)-mode bipartite-graph network, the Hamiltonian can be expressed as 
\begin{equation}
	\hat{H}_{\text{cbn}}=(\hat{a}^{\dagger}_{1},\hat{a}^{\dagger}_{2},\hat{B}_{1}^{\dagger})\textbf{H}^{(1)}_{aB}\left(\begin{array}{c}
		\hat{a}_{1}	\\
	\hat{a}_{2}	\\
		\hat{B}_{1}	\\
	\end{array}\right),
\end{equation}
where the coefficient matrix is given by
\begin{equation}
	\textbf{H}_{aB}^{(1)}=\left(\begin{array}{cc|c}
		\omega_{\text{1}} & g & G_{1,1} \\
		g^{*} & \omega_{\text{2}}& G_{2,1} \\
		\hline G_{1,1}^{*} &G_{2,1}^{*}  & \Omega_{1}
	\end{array}\right).~\label{HaB21}
\end{equation}
We can see from Eq.~(\ref{HaB21}) that there are three connections in the $(2+1)$-mode bipartite-graph network. In practice, both the direct sender-receiver coupling and the indirect coupling through the normal mode $B_{1}$ are crucial for the transfer of excitations from the sender to the receiver, so it is an important topic to analyze the influence of these couplings on the excitation transfer.

For the ($2+1$)-mode network, there exist four different coupling configurations according to the distribution of these connections, as shown in Fig.~\ref{mode_2_1}(a).  To focus on the influence of the normal mode $B_{1}$ on the excitation transfer, below we consider the case where the frequencies of  the sender and receiver are degenerate, i.e., $\omega_{1}=\omega_{2}=\omega_{m}$. Since the detuning between the modes $\{a_{1}$, $a_{2}\}$ and the intermediate mode $B_{1}$ will affect the excitation transfer efficiency, we will also investigate the dependence of the excitation transfer on the detuning. 

In the following, we study the excitation transfer in these four coupling configurations. When these three coupling strengths $g$, $G_{1,1}$, and $G_{2,1}$ exist, we assume that the coupling strengths are real. In the case $G_{1,1}=G_{2,1}=G$,  we can obtain the mean excitation number of the receiver mode $a_{2}$ based on Eq.~(\ref{a2daga2}),
\begin{eqnarray}
	\langle \hat{a}_{2}^{\dagger}\hat{a}_{2}(t)\rangle  & = & -\frac{\Omega_{R}\cos\left(\Omega_{R}t/2\right)\cos\left[(3g-\Delta)t/2\right]}{2\Omega_{R}}\notag\\  & &+\frac{(g+\Delta)\sin\left(\Omega_{R}t/2\right)\sin\left[(3g-\Delta)t/2\right]}{2\Omega_{R}}\notag\\
	&  & +\frac{2G^{2}\left[\cos(\Omega_{R}t)-1\right]+\Omega_{R}^{2}}{2\Omega_{R}^{2}},~\label{a2daga2lambda}
\end{eqnarray}
where $\Omega_{R}=\sqrt{8G^{2}+(g+\Delta)^{2}}$ with $\Delta=\omega_{m}-\Omega_{1}$. We can see from Eq.~(\ref{a2daga2lambda}) that the mean excitation number depends on the detuning $\Delta$ and the coupling strengths $g$ and $G$. Therefore, the excitation transfer efficiency can be evaluated by calculating the maximum mean excitation number $\langle \hat{a}_{2}^{\dagger}\hat{a}_{2}\rangle_{\text{max}}$ for the receiver mode $a_{2}$ corresponding to these four configurations described in Fig.~\ref{mode_2_1}(a). 

In Figs.~\ref{mode_2_1}(b) and \ref{mode_2_1}(c), we plot the maximum value $\langle \hat{a}_{2}^{\dagger}\hat{a}_{2}\rangle_{\text{max}}$ as a function of the coupling strength $G$ and the detuning $\Delta$, corresponding to the presence and absence of the direct coupling $g$, respectively, when the normal mode $B_{1}$ is commonly coupled to both the sender mode $a_{1}$ and the receiver mode $a_{2}$. We can see from Fig.~\ref{mode_2_1}(b) that the single excitation is near perfectly transferred in most cases, and that the value of $\langle \hat{a}_{2}^{\dagger}\hat{a}_{2}\rangle_{\text{max}}$ is smaller than 1 for some parameters. For example, based on Eq.~(\ref{a2daga2lambda}), the maximum value  $\langle \hat{a}^{\dagger}_{2}\hat{a}_{2}\rangle_{\text{max}}$ satisfies $	\langle \hat{a}^{\dagger}_{2}\hat{a}_{2}\rangle_{\text{max}}=4g^{2}/(3g-\Delta)^{2}<1$ when $G=\sqrt{g(g-\Delta)}$ for $\Delta< g$. When the connection between the sender and the receiver is cut off, i.e., $g=0$, the result is shown in Fig.~\ref{mode_2_1}(c). We find that the maximum value $\langle \hat{a}^{\dagger}_{2}\hat{a}_{2}\rangle_{\text{max}}=1$,  for $G\neq \vert \Delta\vert$; while  $\langle \hat{a}^{\dagger}_{2}\hat{a}_{2}\rangle_{\text{max}}=0.75$ for $G=\vert \Delta\vert$. 
In addition, when the normal mode $B_{1}$ is only coupled to either the sender mode $a_{1}$ or the receiver mode $a_{2}$, i.e., $G_{2,1}=0$ or $G_{1,1}=0$, we show $\langle \hat{a}_{2}^{\dagger}\hat{a}_{2}\rangle_{\text{max}}$ as a function of $G$ and $\Delta$ in Figs.~\ref{mode_2_1}(d) and~\ref{mode_2_1}(e). We can see from Figs.~\ref{mode_2_1}(d) and~\ref{mode_2_1}(e) that the value of $\langle \hat{a}_{2}^{\dagger}\hat{a}_{2}\rangle_{\text{max}}$ decreases as $G$ increases, while it can reach 1 at $G/\vert\Delta\vert\ll 1$. The decay rate takes the largest value in the resonance case, i.e., $\Delta=0$. This is because the excitation transfer between the mode $a_{1}$ ($a_{2}$) and the intermediate normal mode $B_{1}$ is faster in the resonance case $\Delta=0$. These results indicate that the normal mode $B_{1}$ coupled to either the sender or the receiver will inhibit the excitation transfer. We point out that the results in these two cases ($G_{21}=0$ and $G_{11}=0$) are the same. 

For the case of  $G_{1,1}\neq G_{2,1}$, we calculate the maximum mean excitation number $\langle \hat{a}_{2}^{\dagger}\hat{a}_{2}\rangle_{\text{max}}$ as a function of the coupling strength $G$ ($G=G_{2,1}$) and the detuning $\Delta$ with $G_{1,1}=0.008\omega_{m}$. The results for these two cases ($g\neq 0$ and $g=0$) are shown in Figs.~\ref{mode_2_1}(f) and \ref{mode_2_1}(g). In Fig.~\ref{mode_2_1}(f), the maximal value $\langle \hat{a}_{2}^{\dagger}\hat{a}_{2}\rangle_{\text{max}}$ decreases as $G$ increases for the near-resonance case when $G\neq 0.008\omega_{m}$, while the highly efficient transfer can be achieved at $G=0.008\omega_{m}$ with $\Delta\neq 0.0036\omega_{m}$.  For $g=0$ [Fig.~\ref{mode_2_1}(g)], the value of $\langle \hat{a}_{2}^{\dagger}\hat{a}_{2}\rangle_{\text{max}}$ increases to unity and then decreases as $G$ increases, and $\langle \hat{a}_{2}^{\dagger}\hat{a}_{2}\rangle_{\text{max}}=1$ at $G=0.008\omega_{m}$ and $\Delta\neq \pm G$. These results indicate that a highly efficient excitation transfer is possible in the near-resonance case only when $G_{1,1}=G_{2,1}$.

\begin{figure*}[tbp]
	\center
	\includegraphics[width=1 \textwidth]{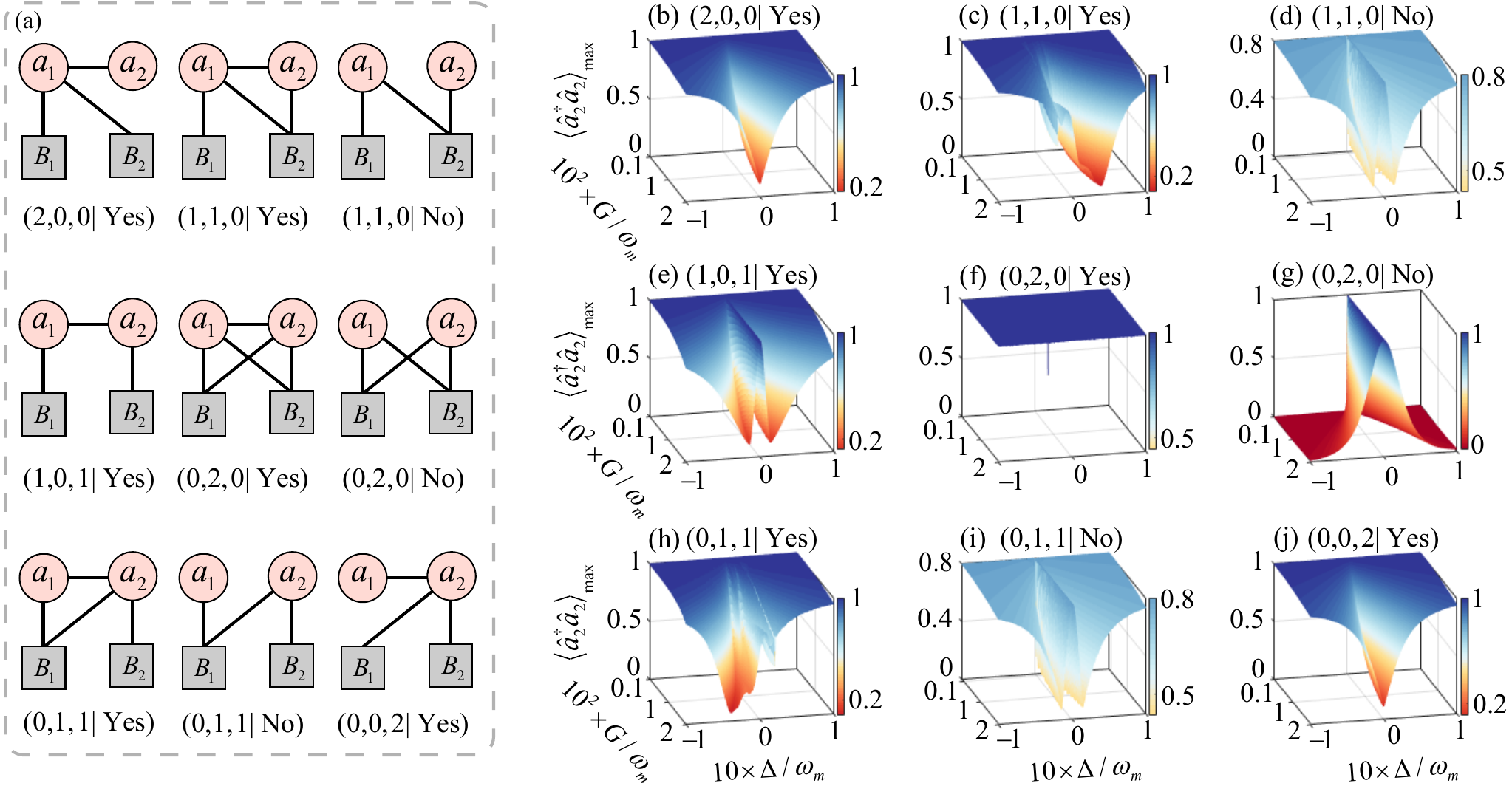}
	\caption{(a) Nine coupling configurations of the ($2+2$)-mode bipartite-graph networks,  where the configurations are labeled by $(N_{1},N_{2},N_{3}|\text{Yes or No})$. [(b)-(j)] Maximum mean excitation number $\langle \hat{a}^{\dagger}_{2}\hat{a}_{2}\rangle_{\text{max}}$ corresponding to these nine configurations depicted in panel (a) vs the coupling strength $G/\omega_{m}$ and the detuning $\Delta/\omega_{m}$. Other  parameters used are $\omega_{1}=\omega_{2}=\omega_{m}$ and $g=0.01\omega_{m}$ when $g$ exists.}
	\label{mode_2_2_mesh}
\end{figure*}

\subsection{Quantum excitation transfer in a ($2+2$)-mode bipartite-graph network}
We now turn to the ($2+2$)-mode bipartite-graph-network case, in which there exist two intermediate normal modes. The Hamiltonian of the ($2+2$)-mode bipartite-graph network can be expressed as
\begin{equation}
		\hat{H}_{\text{cbn}}=(\hat{a}^{\dagger}_{1},\hat{a}^{\dagger}_{2},\hat{B}_{1}^{\dagger},\hat{B}_{2}^{\dagger})\textbf{H}^{(2)}_{aB}\left(\begin{array}{c}
		\hat{a}_{1}	\\
		\hat{a}_{2}	\\
	\hat{B}_{1}	\\
		\hat{B}_{2}\\
	\end{array}\right),
\end{equation}
where the coefficient matrix reads
\begin{equation}
	\textbf{H}_{aB}^{(2)}=\left(\begin{array}{cc|cc}
		\omega_{\text{1}} & g & G_{1,1}& G_{1,2} \\
		g^{*} & \omega_{\text{2}}& G_{2,1}& G_{2,2}  \\
		\hline G_{1,1}^{*} &G_{2,1}^{*}  & \Omega_{1}& 0\\
		G_{1,2}^{*} &G_{2,2}^{*}& 0  & \Omega_{2}\\
	\end{array}\right),~\label{HaB1B2}
\end{equation}
with $\Omega_{1}=\omega_{m}-\Delta$ and $\Omega_{2}=\omega_{m}+\Delta$.
To analyze the transfer of a single excitation from the sender mode $a_{1}$ to the receiver mode $a_{2}$ in the (2+2)-mode bipartite-graph network, we consider different coupling configurations by controlling the coupling channels $g$ and $G_{k,j}$ with $k,j=1,2$. For convenience, we assume that both the sender and receiver modes have the same resonance frequency, i.e., $\omega_{1}=\omega_{2}=\omega_{m}$. Similarly, we now assume for simplicity that, if the couplings associated with the normal modes are present, all coupling strengths are equal, i.e., $G_{k,l}=G$. To ensure that the sender and the receiver are directly or indirectly coupled, there exist nine coupling configurations, as shown in Fig.~\ref{mode_2_2_mesh}(a).

Based on these nine coupling configurations, we can analyze the transfer of a single excitation from the sender to the receiver, and we also show the maximum value $\langle \hat{a}_{2}^{\dagger}\hat{a}_{2}\rangle_{\text{max}}$ of the mean excitation number as a function of the coupling strength $G$ and the detuning $\Delta$ in Figs.~\ref{mode_2_2_mesh}(b-j). We can see from Fig.~\ref{mode_2_2_mesh}(b) that the $\langle \hat{a}_{2}^{\dagger}\hat{a}_{2}\rangle_{\text{max}}$ decreases with the increasing $G$ when the frequencies of the two normal modes and the sender mode are nearly degenerate, i.e., $\Delta\approx 0$. This indicates that the normal modes coupled only to the sender will degrade the excitation transfer efficiency from the sender to the receiver. 

From Fig.~\ref{mode_2_2_mesh}(c), we find that $\langle \hat{a}_{2}^{\dagger}\hat{a}_{2}\rangle_{\text{max}}$ decreases when increasing $G$ in the case of $G_{2,1}=0$ and $g\neq 0$, and the decay rate depends on the detuning $\Delta$. For $g=0$, we see from Fig.~\ref{mode_2_2_mesh}(d) that the maximum value $\langle \hat{a}_{2}^{\dagger}\hat{a}_{2}\rangle_{\text{max}}$ only reaches 0.8. These results indicate that the transfer efficiency of a single excitation from the sender to the receiver can be suppressed when the normal mode $B_{1}$ is only coupled to the sender. 

\begin{figure*}[tbp]
	\center
	\includegraphics[width=0.95 \textwidth]{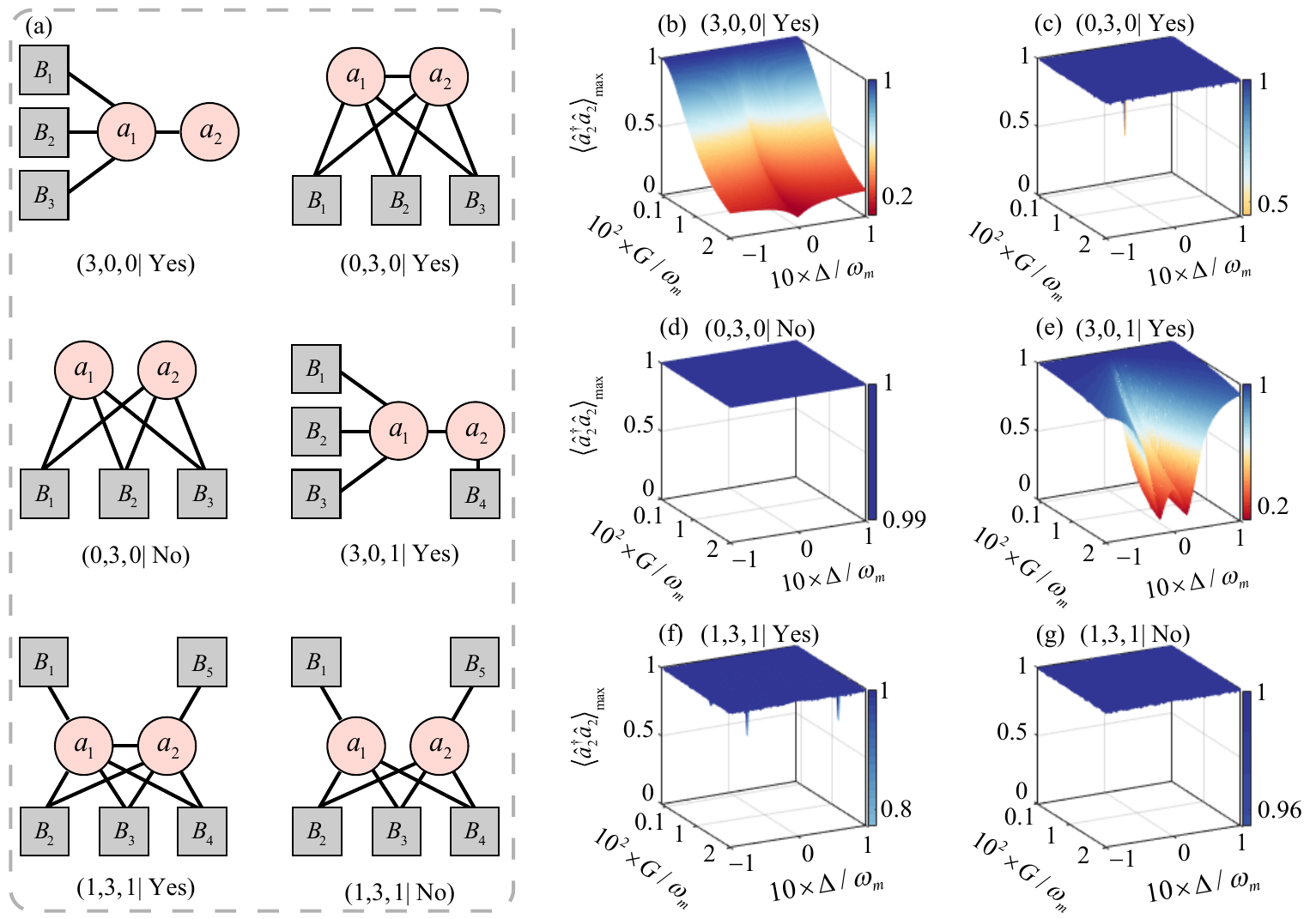}
	\caption{(a) Six coupling configurations of the ($2+N_{\text{c}}$)-mode bipartite-graph networks with $N_{\text{c}}=3,4,5$, where the configurations are labeled by $(N_{1},N_{2},N_{3}|\text{Yes or No})$. [(b)-(g)] Maximum mean excitation number $\langle \hat{a}^{\dagger}_{2}\hat{a}_{2}\rangle_{\text{max}}$ as a function of the coupling strength $G/\omega_{m}$ and the detuning $\Delta/\omega_{m}$ corresponding to these six configurations depicted in panel (a). Here, we consider all real coupling strengths $G_{k,j}$ to be equal. The frequencies of these normal modes are $\Omega_{1}=\omega_{m}-\Delta$, $\Omega_{2}=\omega_{m}$,  $\Omega_{3}=\omega_{m}+\Delta$ in panels (b-d), $\Omega_{1}=\omega_{m}-\Delta$, $\Omega_{2}=\Omega_{4}=\omega_{m}$,  $\Omega_{3}=\omega_{m}+\Delta$ in panel (e), and $\Omega_{2}=\omega_{m}-\Delta$, $\Omega_{4}=\omega_{m}+\Delta$, $\Omega_{1}=\Omega_{3}=\Omega_{5}=\omega_{m}$  in panels (f-g). Other parameters used are $\omega_{1}=\omega_{2}=\omega_{m}$.}	
	\label{three_dao_five}
\end{figure*}

When both the sender and the receiver are coupled to an individual normal mode, the maximum value $\langle \hat{a}_{2}^{\dagger}\hat{a}_{2}\rangle_{\text{max}}$ as a function of $G$ and $\Delta$ is shown in Fig.~\ref{mode_2_2_mesh}(e). We find that the transfer efficiency can reach 1 for any $G$ in the case $\Delta=0$, while it decreases as $G$ increases for $\Omega_{1}\neq \Omega_{2}$. 
When $g\neq 0$ and the two normal modes are coupled to both the sender and the receiver, we see from Fig.~\ref{mode_2_2_mesh}(f) that the maximum value $\langle \hat{a}_{2}^{\dagger}\hat{a}_{2}\rangle_{\text{max}}$ is nearly 1 when the frequencies of the two normal modes are nondegenerate, i.e., $\Omega_{1}\neq\Omega_{2}$. However, the maximum value $\langle \hat{a}_{2}^{\dagger}\hat{a}_{2}\rangle_{\text{max}}$ is only 4/9 at $G=g/\sqrt{2}$ when $\Omega_{1}=\Omega_{2}=\omega_{m}$. This is because there is a dark mode decoupled from the sender mode $a_{1}$ that suppresses the transfer efficiency, and the dark mode can be expressed as $B_{-}=(B_{1}+B_{2}-\sqrt{2}a_{2})/2$. 
When the sender and the receiver are not directly coupled, i.e., $g=0$, we find from Fig.~\ref{mode_2_2_mesh}(g) that $\langle \hat{a}_{2}^{\dagger}\hat{a}_{2}\rangle_{\text{max}}$ decreases from 1 to 0 as $\vert\Delta\vert$ increases. In this case, the transfer efficiency can always reach 1 under proper parameters. 

When the normal mode $B_{2}$ is only coupled to the receiver, the result for $g\neq 0$ ($g= 0$) is shown in Fig.~\ref{mode_2_2_mesh}(h) [Fig.~\ref{mode_2_2_mesh}(i)]. We find that the results are similar to those in Figs.~\ref{mode_2_2_mesh}(c) and \ref{mode_2_2_mesh}(d). Meanwhile, the result in Fig.~\ref{mode_2_2_mesh}(j) is the same as that in Fig.~\ref{mode_2_2_mesh}(b) when the two normal modes are only coupled to the receiver. This phenomenon indicates that the normal modes only coupled to either the sender or the receiver have the same effect on the excitation transfer. 

Based on the above analyses concerning the excitation transfer in the (2+2)-mode bipartite-graph network, we can deduce the following points: (i) The transfer efficiency decreases with the increase of the coupling strength $G$ when there is at least one normal mode coupled only to either the sender or the receiver; (ii) The transfer efficiency can reach 1 for $g\neq \sqrt{2}G$ at $\Delta=0$  when the normal modes are jointly coupled to both the sender and the receiver; (iii) The transfer efficiency can reach 1 for any $G$ in the case of $\Delta=0$ for the (2+2)-mode symmetric chain network. Based on these analyses, we can infer the parameter conditions for highly efficient excitation transfer in the (2+2)-mode bosonic network.

\begin{table*}[t]
	\footnotesize
	\caption{Summary of the transfer efficiency for various network configurations within the single-excitation subspace. 
		“Always’’ (“Never’’) indicates that a highly efficient (close to 1) transfer is always (never) possible, 
		while “Conditional’’ indicates that a highly efficient transfer requires certain conditions of the detuning $\Delta$ and the coupling strengths $G$ and $g$. 
		The frequencies of $a_{1}$ and $a_{2}$ used in our simulation
		are resonant ($\omega_{1}=\omega_{2}=\omega_{m}$) and all coupling strengths between the sender or receiver mode and the normal modes $B_{1\text{-}N_{\text{c}}}$ are identical ($G_{1,j}=G_{2,j}=G$ for $j=1,2,\ldots,N_{\text{c}}$). The notation $(N_1,N_2,N_3|\text{Yes or No})$ with $N_{\text{c}}=N_1+N_2+N_3$ is introduced in Fig.~\ref{system_model}.}~\label{tab1}	
	\tabcolsep 1pt
	\begin{tabular*}{1\textwidth}{@{\extracolsep{\fill}}ccc|ccc|ccc}
		\toprule
		\multicolumn{3}{c|}{$N_{\text{c}}=0, 1$} &\multicolumn{3}{c|}{$N_{\text{c}}=2$}&\multicolumn{3}{c}{$N_{\text{c}}=3, 4, 5$} \\ \hline
		Configurations & Highly efficient transfer & Conditions &Configurations&Highly efficient transfer &Conditions&Configurations & Highly efficient transfer & Conditions\\
		\hline $(0,0,0|\text{Yes})$ & Always & --&$(2,0,0|\text{Yes})$&Conditional &$G\ll \vert\Delta\vert$& $(3,0,0|\text{Yes})$&Conditional&$G\ll \vert\Delta\vert$ \\
		$(0,1,0|\text{Yes})$ & Conditional &$G\neq \sqrt{g(g-\Delta)}$ &$(1,1,0|\text{Yes})$&Conditional &$G\ll \vert\Delta\vert$& $(0,3,0|\text{Yes})$&Conditional &$G\neq g/\sqrt{3}$\\
		$(0,1,0|\text{No})$ & Conditional&$G\neq \pm \Delta$&$(1,1,0|\text{No})$&Never&--& $(0,3,0|\text{No})$&Always&-- \\
		$(1,0,0|\text{Yes})$ & Conditional & $G\ll \vert\Delta\vert$&$(1,0,1|\text{Yes})$&Conditional&$\Delta\approx 0$& $(3,0,1|\text{Yes})$& Conditional&$G\ll \vert\Delta\vert$ \\
		$(0,0,1|\text{Yes})$ & Conditional & $G\ll \vert\Delta\vert$&$(0,2,0|\text{Yes})$&Conditional &$G\neq g/\sqrt{2}$&  $(1,3,1|\text{Yes})$&Conditional&$G\ll \vert\Delta\vert$\\
		& & &$(0,2,0|\text{No})$&Conditional&$\Delta \approx 0$ &$(1,3,1|\text{No})$&Always&--\\
		& & &$(0,1,1|\text{Yes})$&Conditional& $G\ll \vert\Delta\vert$&&&\\
		& & &$(0,1,1|\text{No})$&Never&-- &&&\\
		& & &$(0,0,2|\text{Yes})$&Conditional& $G\ll \vert\Delta\vert$&&&\\
		\botrule
	\end{tabular*}
\end{table*}

\subsection{Quantum excitation transfer in a ($2+N$)-mode bipartite-graph network for $N_{\text{c}}=3,4,5$.}
To further study quantum excitation transfer in the bosonic networks under the bipartite-graph framework, we consider $N_{\text{c}}=3,4,5$ for the ($2+N_{\text{c}}$)-mode bipartite-graph network as examples. For our purposes,  we now consider a general ($2+5$)-mode bipartite-graph network, which can be reduced to $N_{\text{c}}=3,4$ by cutting off some connections. The Hamiltonian of the ($2+5$)-mode network can be expressed as 
\begin{equation}
		\hat{H}_{\text{cbn}}=(\hat{a}^{\dagger}_{1},\hat{a}^{\dagger}_{2},\hat{B}_{1}^{\dagger},\hat{B}_{2}^{\dagger},\ldots,\hat{B}_{5}^{\dagger})\textbf{H}^{(5)}_{aB}\left(\begin{array}{c}
		\hat{a}_{1}	\\
		\hat{a}_{2}	\\
		\hat{B}_{1}	\\
		\hat{B}_{2}\\
		\vdots\\
		\hat{B}_{5}
	\end{array}\right),
\end{equation}
with the coefficient matrix
\begin{equation}
	\textbf{H}_{aB}^{(5)}=\left(\begin{array}{cc|ccccc}
		\omega_{\text{1}} & g & G_{1,1}& G_{1,2}& G_{1,3} & G_{1,4} & G_{1,5}  \\
		g^{*} & \omega_{\text{2}}& G_{2,1}& G_{2,2} & G_{2,3} & G_{2,4} & G_{2,5}  \\
		\hline G_{1,1}^{*} &G_{2,1}^{*}  & \Omega_{1}& 0& 0& 0& 0\\
		G_{1,2}^{*} &G_{2,2}^{*}& 0  & \Omega_{2}& 0& 0& 0\\
		G_{1,3}^{*} &G_{2,3}^{*}& 0  & 0& \Omega_{3}& 0& 0\\
		G_{1,4}^{*} &G_{2,4}^{*}& 0  & 0& 0& \Omega_{4}& 0\\
		G_{1,5}^{*} &G_{2,5}^{*}& 0  & 0& 0& 0& \Omega_{5}
	\end{array}\right).~\label{HaB1B2B3B4B5}
\end{equation}
We can find from Eq.~(\ref{HaB1B2B3B4B5}) that there are 36 coupling configurations by controlling the coupling strengths $g$, $G_{k,j}$ with $k=1,2$ and $j=1,2,\ldots,5$. Here, we only show the results for six typical configurations as examples in Fig.~\ref{three_dao_five}(a). For convenience, we denote these configurations as $(N_{1},N_{2},N_{3}|\text{Yes or No})$. Here, $N_{\text{c}}=N_{1}+N_{2}+N_{3}$, and the notation ``Yes" (``No'') means that the sender and the receiver are directly connected (disconnected). Note that, in our simulations, we consider that there exists a frequency distribution for these intermediate modes corresponding to $N_{1},N_{2}$, and $N_{3}$, when $N_{1},N_{2},N_{3}>1$.

Based on Eq.~(\ref{a2daga2}), we numerically calculate the  mean excitation number $\langle \hat{a}_{2}^{\dagger}\hat{a}_{2}(t)\rangle$, and show the maximum value $\langle \hat{a}_{2}^{\dagger}\hat{a}_{2}\rangle_{\text{max}}$ as a function of $G/\omega_{m}$ and $\Delta/\omega_{m}$ [see Figs.~\ref{three_dao_five}(b-g)]. Here, we consider that all the coupling strengths $G_{k,j}$ are real and equal, i.e., $G_{k,j}=G$. We see from Fig.~\ref{three_dao_five}(b) that the maximum value $\langle \hat{a}^{\dagger}_{2}\hat{a}_{2}\rangle_{\text{max}}$ is symmetric with respect to $\Delta=0$, and it decreases as $G$ increases in the case of  $(3,0,0|\text{Yes})$, which means that the transfer efficiency can be suppressed when these three normal modes are only coupled to the sender. In the two cases $(0,3,0|\text{Yes})$ and $(0,3,0|\text{No})$, both the sender and the receiver are commonly coupled with three normal modes, the numerical results are shown in Figs.~\ref{three_dao_five}(c) and~\ref{three_dao_five}(d). Figure~\ref{three_dao_five}(c) indicates that $\langle \hat{a}_{2}^{\dagger}\hat{a}_{2}\rangle_{\text{max}}\approx 1$ at $G\neq g/\sqrt{3}$; while  $\langle \hat{a}_{2}^{\dagger}\hat{a}_{2}\rangle_{\text{max}}=4/9$ when $G=g/\sqrt{3}$ in the degenerate case, i.e., $\Delta=0$. This is because there are three dark modes with respect to the sender mode $a_{1}$ in this case, and hence the excitation cannot be completely transferred from the sender to the receiver. 

When $g=0$, we see from Fig.~\ref{three_dao_five}(d) that the $\langle \hat{a}_{2}^{\dagger}\hat{a}_{2}\rangle_{\text{max}}$ reaches 1, which means that the excitation in the sender can be completely transferred to the receiver through the paths $a_{1}\rightarrow B_{j=\text{1,2,3}}\rightarrow a_{2}$  in the case of $(0,3,0|\text{No})$. When these three normal modes $B_{1,2,3}$ are coupled to the sender and the single normal mode $B_{4}$ is coupled to the receiver, we show the $\langle \hat{a}_{2}^{\dagger}\hat{a}_{2}\rangle_{\text{max}}$ as a function of the coupling strength $G$ and the detuning $\Delta$ in Fig.~\ref{three_dao_five}(e). We can find that the value of $\langle \hat{a}_{2}^{\dagger}\hat{a}_{2}\rangle_{\text{max}}$ decreases when increasing $G$, which means that the transfer efficiency is suppressed in this case. 

For the two cases  $(1,3,1|\text{Yes})$ and $(1,3,1|\text{No})$, we also numerically simulate $\langle \hat{a}_{2}^{\dagger}\hat{a}_{2}\rangle_{\text{max}}$ as a function of $G/\omega_{m}$ and $\Delta/\omega_{m}$ in Figs.~\ref{three_dao_five}(f) and~\ref{three_dao_five}(g), respectively. We find that $\langle \hat{a}_{2}^{\dagger}\hat{a}_{2}\rangle_{\text{max}}=0.7713$ at $G/\omega_{m}=0.0185$ and $\vert\Delta\vert/\omega_{m}=0.0675$ for $g/\omega_{m}= 0.01$, which means that these transmission paths exhibit a destructive interference under this parameter condition and the single excitation transfer efficiency from the sender to the receiver is suppressed. When $g=0$, we find from Fig.~\ref{three_dao_five}(g) that $\langle \hat{a}_{2}^{\dagger}\hat{a}_{2}\rangle_{\text{max}}$ approaches 1. 

Based on the above analyses, we can see that the bipartite-graph framework is convenient for studying quantum excitation transfer in a bosonic network with finite normal modes. It can be used to simplify the network configuration and to optimize the analyses of  the transfer paths, thereby giving the parameter conditions for a highly efficient quantum transfer. Crucially, this framework also facilitates the elucidation of the generation mechanism of dark modes. In the bipartite-graph representation, a dark mode appears when a linear superposition of intermediate normal modes exhibits zero effective coupling to both the sender and receiver modes~\cite{Jian2023}. Consequently, the excitations within such modes become trapped and cannot reach the receiver mode. To identify the parameter conditions for achieving an efficient transfer in the cases of $N_{\text{c}}=1,2,\ldots,5$, we introduce Table~\ref{tab1}. Furthermore, using the transformation relation in Eq.~(\ref{BjU}), the parameter conditions corresponding to a highly efficient quantum transfer in the original bosonic network can be obtained accordingly.

\subsection{Excitation transfer efficiency for the $(2+N_{\text{c}})$-mode network with a finite $N_{\text{c}}$ under the dissipations}
In the open system case, the networks inevitably suffers from the dissipations due to the couplings of the system with its environment. As a result, the dynamical evolution of the network becomes nonunitary, leading to an energy loss and reduced excitation transfer efficiency. To capture the dissipations, we assume that each node on the network is coupled to its own local vacuum environment.

To quantitatively evaluate the performance of the excitation transfer under the dissipations, we calculate the cumulative output~\cite{Plenio2008,Caruso2009} from the receiver mode into its local vacuum environment. Physically, the excitations dissipated into the environment of the receiver mode should be transferred from the sender mode to the receiver mode in advance. Therefore, the total number of the excitations emitted from the receiver provides a natural and faithful measure of the transfer efficiency in the presence of dissipations.
Concretely, we adopt the input-output relation~\cite{Gardiner2004,Walls2008}
\begin{equation}
	o_{\text{in}}(t)+o_{\text{out}}(t)=\sqrt{\kappa}o(t)~\label{inputoutput}
\end{equation}
for a bosonic mode, where the input field operator $o_{\text{in}}(t)$ and the output field operator $o_{\text{out}}(t)$ corresponding to the bosonic mode are given by
\begin{subequations}
	\begin{align}
	o_{\text{in}}(t)&=-\frac{1}{\sqrt{2\pi}}\int_{-\infty}^{+\infty}d\omega e^{-i\omega(t-t_{0})}c(\omega,t_{0}),\\
	o_{\text{out}}(t)&=\frac{1}{\sqrt{2\pi}}\int_{-\infty}^{+\infty}d\omega e^{-i\omega(t-t_{1})}c(\omega,t_{1}).
	\end{align}
\end{subequations}
Here, $c(\omega,t_{0})$ and $c(\omega,t_{1})$ are the annihilation operators of the continuous modes for its environment at the initial time $t_{0}$ and the final time $t_{1}$, respectively. They satisfy the standard bosonic commutative relation $[c(\omega,t_{i}),c^{\dagger}(\omega^{\prime},t_{i})]=\delta(\omega-\omega^{\prime})$ for $i=0,1$.

To calculate the cumulative output of the receiver mode $a_{2}$, we assume that the initial field is a vacuum. Based on Eq.~(\ref{inputoutput}), the cumulative output of the receiver mode $a_{2}$ is given by
\begin{equation}
	\eta_{a_{2},\text{out}}=\kappa_{2}\int_{0}^{\infty} \langle a_{2}^{\dagger}a_{2}(t)\rangle dt,
\end{equation}
where the mean excitation number $\langle a_{2}^{\dagger}a_{2}(t)\rangle$ is given in Eq.~(\ref{a2daga2}).
For the present system, there exists
the conservation law 
\begin{equation}
	\eta_{a_{2},\text{out}}+\eta_{a_{1},\text{out}}+\sum_{j=1}^{N_{2}}\eta_{B_{j},\text{out}}=1,
\end{equation}
where $\eta_{a_{1},\text{out}}$ and $\eta_{B_{j},\text{out}}$ are the cumulative output of the sender mode $a_{1}$ and the $j$th normal mode $B_{j}$, respectively.

\begin{figure}[tbp]
	\center
	\includegraphics[width=0.45 \textwidth]{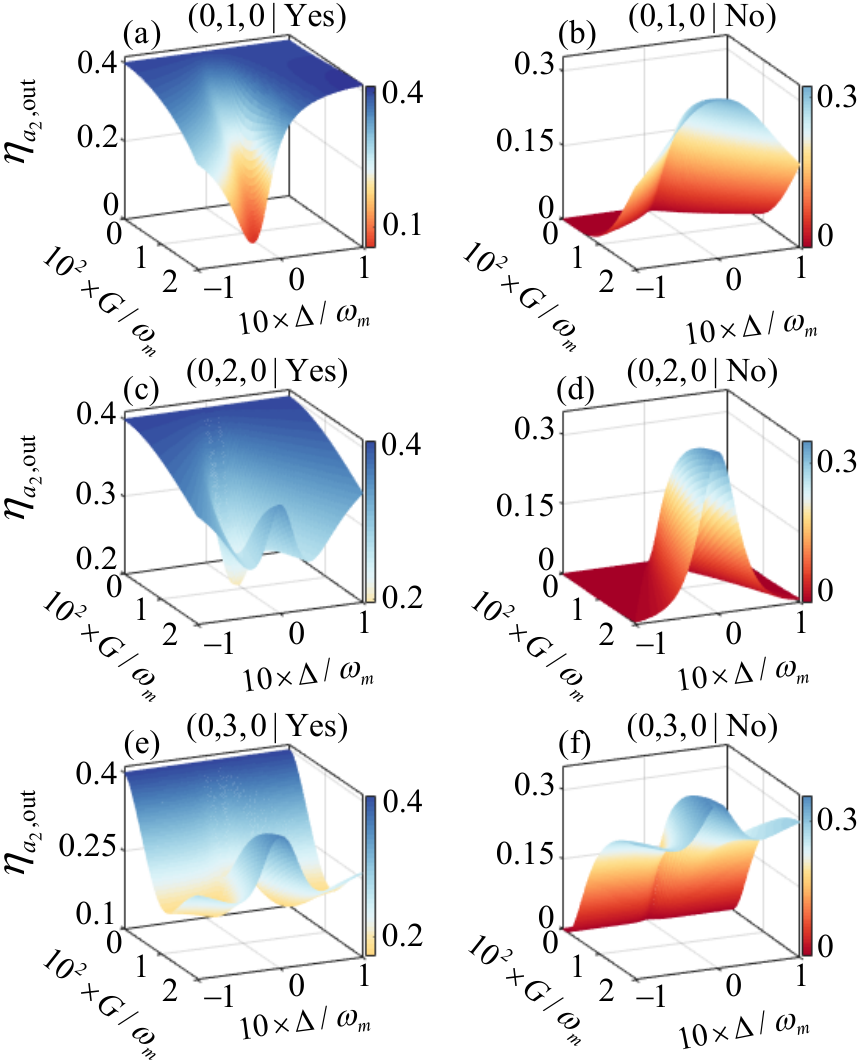}
	\caption{Cumulative output $\eta_{a_{2},\text{out}}$ of the receiver mode $a_{2}$ as a function of the coupling strength $G/\omega_{m}$ and the detuning $\Delta/\omega_{m}$ for the coupling configurations $(0,N_{2},0|\text{Yes or No})$ with $N_{2}=1,2,3$. The decay rates are [(a) and (b)] $\gamma_{1}=0.01\omega_{m}$, [(c) and (d)] $\gamma_{1}=\gamma_{2}=0.01\omega_{m}$, and [(e) and (f)] $\gamma_{1}=\gamma_{2}=\gamma_{3}=0.01\omega_{m}$. Other parameters used are $\omega_{1}=\omega_{2}=\omega_{m}$, $\kappa_{1}=\kappa_{2}=0.01\omega_{m}$,  and $g=0.01\omega_{m}$ when $g$ exists.}
	\label{ott}
\end{figure}

As examples, we consider the case where the normal modes are coupled simultaneously to both the sender mode $a_{1}$ and the receiver mode $a_{2}$, i.e., the configuration $(0,N_{2},0|\text{Yes or No})$. We examine these six cases by choosing $N_{2}=1,2,3$. For convenience, we consider that all the decay rates are identical, i.e., $\kappa_{1}=\kappa_{2}=\gamma_{1\text{-}N_{2}}=\kappa$, so that these noise operators in Eq.~(\ref{tilNt}) can be expressed as $F_{o}(t)=\sqrt{\kappa}o_{\text{in}}(t)$ with $o=\{a_{1},a_{2},B_{1\text{-}N_{2}}\}$. In this case, we display the cumulative output $\eta_{a_{2},\text{out}}$ as the coupling strength $G/\omega_{m}$ and the detuning $\Delta/\omega_{m}$ in Fig.~\ref{ott}.
	
For the case of $N_{2}=1$, when $g\neq 0$ [Fig.~\ref{ott}(a)], the cumulative output
$\eta_{a_{2},\text{out}}$ decreases as the coupling strength 
$G$ increases at $\Delta<g$. The strongest suppression occurs under the condition $G=\sqrt{g(g-\Delta)}$, in consistent with the results in the closed-system case. In contrast, when the direct coupling vanishes, i.e., $g=0$, we can see from Fig.~\ref{ott}(b) that the cumulative output increases as the coupling strength 
$G$ increases, with the fastest growth occurring at resonance $\Delta=0$.

When the two normal modes are simultaneously coupled to the sender mode and the receiver mode, the output behaviors become more intricate. For the case of $g\neq 0$ [Fig.~\ref{ott}(c)], the cumulative output $\eta_{a_{2},\text{out}}$ exhibits a nonmonotonic dependence on the coupling strength $G$ near the resonance. It first decreases and then increases as $G$ grows. The minimum output appears precisely at the parameter condition under which a dark mode emerges. This is because the initial excitations trapped in the dark mode will leak into the environment through the dissipation channel of the sender mode $a_{1}$, thereby reducing the transfer efficiency to the receiver mode $a_{2}$. In analogy to Fig.~\ref{ott}(b) for $g=0$, we see from Fig.~\ref{ott}(d) that the cumulative output exhibits a monotonically increasing trend as the coupling strength $G$ increases and reaches its maximum at resonance $\Delta=0$. 

Meanwhile, we also analyze the case where these three normal modes are commonly coupled to the sender mode and the receiver mode. When a direct coupling is present, we can see from Fig.~\ref{ott}(e) that the cumulative output $\eta_{a_{2},\text{out}}$ first decreases and then increases as the coupling strength $G$ increases. Notably, the dip under resonance occurs at the parameter condition under which a dark mode emerges. We also find from Fig.~\ref{ott}(f) that the cumulative output increases monotonically as $G$ increases and reaches its maximum at resonance for $g=0$. Note that when these normal modes are commonly coupled to the sender mode and the receiver mode, i.e., with the coupling configuration $(0,N_{2},0|\text{Yes or No})$, a highly efficient excitation transfer from the sender mode $a_{1}$ to the receiver mode $a_{2}$ occurs at $G/\Delta\ll1$ for $g\neq 0$, while  it only occurs at resonance $\Delta=0$ for $g = 0$.

\begin{figure*}[tbp]
	\center
	\includegraphics[width=0.98 \textwidth]{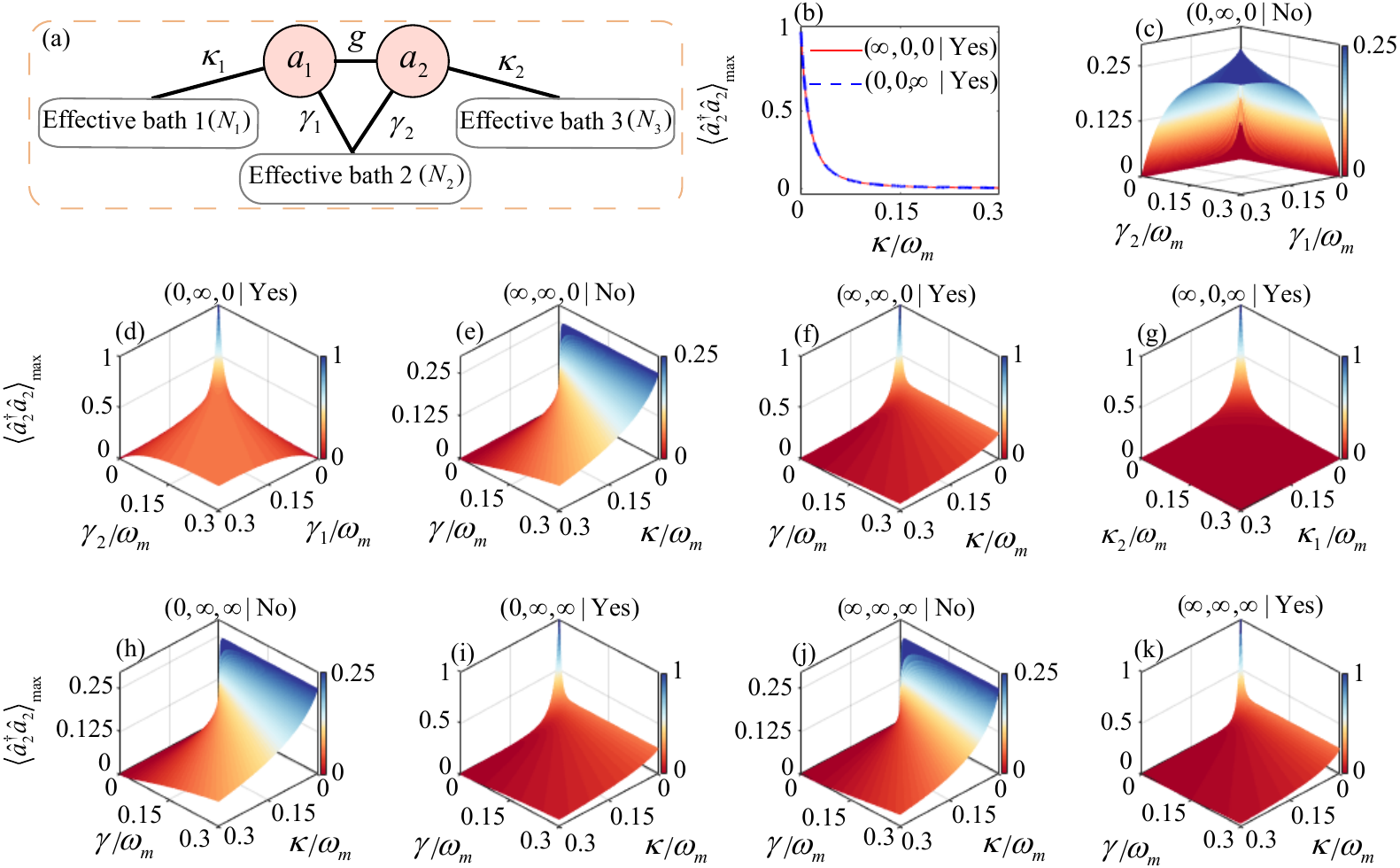}
	\caption{(a) Schematic of the sender and the receiver coupled to three different effective baths. (b) The maximum mean excitation number $\langle \hat{a}^{\dagger}_{2}\hat{a}_{2}\rangle_{\text{max}}$ vs  the effective decay rate $\kappa$ in two cases: $\kappa_{1}=\kappa, \kappa_{2}=\gamma_{1,2}=0, g=0.01\omega_{m}$ (red solid curve) and $\kappa_{2}=\kappa, \kappa_{1}=\gamma_{1,2}=0, g=0.01\omega_{m}$ (blue dashed curve).
		[(c)-(k)] The maximum mean excitation number $\langle \hat{a}^{\dagger}_{2}\hat{a}_{2}\rangle_{\text{max}}$  vs the effective decay rates $\{\kappa_{1,2},\gamma_{1,2}\}$ in these nine cases: (c) $\kappa_{1,2}=g=0$; (d) $\kappa_{1,2}=0, g=0.01\omega_{m}$; (e) $\kappa_{2}=g=0, \kappa_{1}=\kappa, \gamma_{1,2}=\gamma$; (f) $\kappa_{2}=0, \kappa_{1}=\kappa, \gamma_{1,2}=\gamma, g=0.01\omega_{m}$; (g) $\gamma_{1,2}=0, g=0.01\omega_{m}$; (h) $\kappa_{1}=g=0, \kappa_{2}=\kappa, \gamma_{1,2}=\gamma$; (i) $\kappa_{1}=0, \kappa_{2}=\kappa, \gamma_{1,2}=\gamma,g=0.01\omega_{m}$; (j) $\kappa_{1,2}=\kappa, \gamma_{1,2}=\gamma, g=0$; and (k) $\kappa_{1,2}=\kappa, \gamma_{1,2}=\gamma, g=0.01\omega_{m}$. Note that the label $(N_{1},N_{2},N_{3}|\text{Yes or No})$ in panels (b)-(k)  indicates the distribution of $N_{\text{c}}=N_{1}+N_{2}+N_{3}$ normal modes and whether there is a connection between the sender and the receiver.  Other commonly used parameters are $\omega_{1}=\omega_{2}=\omega_{m}$.}
	\label{N1_N2_N3_inf_v1}
\end{figure*}
\section{Quantum excitation transfer in the bipartite-graph network with infinite $N_{\text{c}}$~\label{stifinite}}
In this section, we investigate the transfer of a single excitation from the sender to the receiver in a ($2+N_{\text{c}}$)-mode bipartite-graph network with an infinite $N_{\text{c}}$ normal modes. In particular, we consider the case where the values of $N_{1},N_{2},N_{3}$ take infinity when $N_{i=1,2,3}$ exist in the coupling configurations of the bipartite-graph network. Namely, $N_{i=1,2,3}=0$   ($\infty$) implies there are no (exist) couplings between the sender/receiver and the intermediate normal modes. We should point out that, in a realistic case, the number of these intermediate modes is very large, but it is still finite. Here, we assume that the number is infinite so that we can treat these intermediate modes as baths, and then the evolution of the system can be described by quantum master equation. 

In the above mentioned cases, the Hamiltonian of the networks is given by Eq.~(\ref{HaBaB}). Here, the sender mode $a_{1}$ and the receiver mode $a_{2}$ are only coupled to $N_{1}$ and $N_{3}$ normal modes, respectively, and they are  commonly coupled to $N_{2}$ normal modes. When $N_{i=1,2,3}$ approach infinite, we assume that these intermediate normal modes  initially in  vacuum can be treated as equivalent vacuum baths with discrete mode frequencies. Thus, the bipartite-graph network can be described by a two-bosonic-mode system coupled to three effective vacuum baths, as shown in Fig.~\ref{N1_N2_N3_inf_v1}(a). Then the quantum transfer process of a single excitation from the sender to the receiver in a bipartite-graph network can be reduced to a single-excitation transfer process between two bosonic modes connected to three vacuum baths.

In the above-mentioned infinite-$N_{\text{c}}$ case, the evolution of the system is determined by the quantum master equation
\begin{eqnarray}
	\frac{d\hat{\rho}(t)}{dt}& =&  -i[	\hat{H}_{0},\hat{\rho}(t)]+\mathcal{D}(\sqrt{\kappa_{1}}\hat{a}_{1})[\hat{\rho}(t)]+\mathcal{D}(\sqrt{\kappa_{2}}\hat{a}_{2})[\hat{\rho}(t)]\notag\\
&&+\mathcal{D}(\sqrt{\gamma_{1}}\hat{a}_{1}+\sqrt{\gamma_{2}}\hat{a}_{2})[\hat{\rho}(t)],\label{rhodot}
\end{eqnarray}
where the Hamiltonian reads
\begin{equation}
	\hat{H}_{0}=\omega_{m}\hat{a}_{1}^{\dagger}\hat{a}_{1}+\omega_{m}\hat{a}_{2}^{\dagger}\hat{a}_{2}+ (g \hat{a}_{1}^{\dagger}\hat{a}_{2}+g^{*}\hat{a}_{2}^{\dagger}\hat{a}_{1}).
\end{equation}
The Lindblad superoperator is defined as 
$\mathcal{D}(\hat{O})[\hat{\rho}(t)]=\hat{O}\hat{\rho}(t) \hat{O}^{\dagger}-\hat{\rho}(t) \hat{O}^{\dagger}\hat{O}/2-\hat{O}^{\dagger}\hat{O}\hat{\rho}(t)/2. $
Here, the variables $\kappa_{1}$, $\gamma_{1}$ ({$\kappa_{2}$, $\gamma_{2}$}) are the decay rates associated with the collective couplings of the sender (receiver) with these $N_{1}, N_{2}$ ($N_{3}, N_{2}$) normal modes, which are defined as
\begin{eqnarray}
	\kappa_{1} =2\pi\varrho_{1}(\omega_{m})G_{1}^{2}(\omega_{m}),\hspace{0.5cm}	\kappa_{2} =2\pi\varrho_{2}(\omega_{m})G_{2}^{2}(\omega_{m}),\nonumber \\
	\gamma_{1} =2\pi\varrho_{3}(\omega_{m})G_{1}^{2}(\omega_{m}),\hspace{0.5cm}		\gamma_{2}  =2\pi\varrho_{3}(\omega_{m})G_{2}^{2}(\omega_{m}). ~\label{kappagamma}
\end{eqnarray}
In Eq.~(\ref{kappagamma}), $\varrho_{1,2,3}$ are the spectral densities of these three effective baths, and $G_{1}$ ($G_{2}$) is the effective coupling strength between these baths and the sender mode $a_{1}$ (receiver mode $a_{2}$). In Eq.~(\ref{rhodot}), the second and third terms describe the dissipations caused by the two effective baths 1 and 3, while the last term depicts the dissipation caused by the common effective bath 2.

We emphasize that the derivation of Eq.~(\ref{rhodot}) relies on the flat-spectral-density approximation, i.e., Weisskopf-Wigner approximation. This approximation is justified by the continuum limit ($N_{i=1,2,3} \to \infty$) and the weak-coupling condition ($G \ll \omega_m$). Under these conditions, the intermediate normal modes form a quasicontinuous band, and the bath correlation functions decay on a timescale much faster than the characteristic evolution time of the sender-receiver subnetwork. Consequently, the spectral densities $\varrho_{1,2,3}(\omega)$ can be treated as locally flat in the vicinity of $\omega_{m}$. However, when the coupling strengths $G_{1,2}$ become comparable to the system frequencies, nonMarkovian memory and recurrence effects may become significant.  Furthermore, for large but finite $N_{i=1,2,3}$, the spectrum of the intermediate modes remains discrete, which may lead to deviations from the flat-spectral-density prediction. In these cases, Eq.~(\ref{rhodot}) is no longer valid. Hence, our subsequent study of quantum excitation transfer in the infinite-normal-mode case is conducted under these conditions.

For the bipartite-graph network depicted in Fig.~\ref{N1_N2_N3_inf_v1}(a), we can investigate the transfer of a single excitation for different coupling configurations by controlling these parameters $\kappa_{1,2},\gamma_{1,2}$, and $g$. Here, we assume that the coupling strength $g$  is real. By disconnecting these couplings, there are eleven coupling configurations. For readability, we label these 11 coupling configurations as [$N_{1},N_{2},N_{3}|\text{Yes or No}$], where the mark ``Yes'' (``No'') denotes that the sender and the receiver are connected (disconnected). 

By introducing the vector
\begin{equation}
	\textbf{M}(t)=[\langle \hat{a}_{1}^{\dagger}\hat{a}_{1}\rangle,\langle \hat{a}_{2}^{\dagger}\hat{a}_{2}\rangle,\langle \hat{a}_{1}^{\dagger}\hat{a}_{2}\rangle,\langle \hat{a}_{2}^{\dagger}\hat{a}_{1}\rangle]^{T},
\end{equation}
the equation of motion can be written as
\begin{equation}
	\frac{d\textbf{M}(t)}{dt}=-\textbf{D}\textbf{M}(t),~\label{Mtdot}
\end{equation}
where the coefficient matrix is introduced as 
\begin{equation}
	\textbf{D}=\left(\begin{array}{cccc}
		\kappa_{1}\!+\!\gamma_{1} & 0 & \frac{1}{2}\sqrt{\gamma_{1}\gamma_{2}}\!+\!ig & \frac{1}{2}\sqrt{\gamma_{1}\gamma_{2}}\!-\!ig\\
		0 & \kappa_{2}\!+\!\gamma_{2} & \frac{1}{2}\sqrt{\gamma_{1}\gamma_{2}}\!-\!ig & \frac{1}{2}\sqrt{\gamma_{1}\gamma_{2}}\!+\!ig\\
		\frac{1}{2}\sqrt{\gamma_{1}\gamma_{2}}\!+\!ig & \frac{1}{2}\sqrt{\gamma_{1}\gamma_{2}}\!-\!ig & \frac{1}{2}\sum_{j=1}^{2}(\kappa_{j}\!+\!\gamma_{j}) & 0\\
		\frac{1}{2}\sqrt{\gamma_{1}\gamma_{2}}\!-\!ig & \frac{1}{2}\sqrt{\gamma_{1}\gamma_{2}}\!+\!ig & 0 & \frac{1}{2}\sum_{j=1}^{2}(\kappa_{j}\!+\!\gamma_{j})
	\end{array}\right).
\end{equation}
By solving Eq.~(\ref{Mtdot}), we can obtain the dynamical evolution of the mean excitation number of the receiver mode $a_{2}$ and its maximum value $\langle \hat{a}_{2}^{\dagger}\hat{a}_{2}\rangle_{\text{max}}$. 

For the case of $N_{1}=N_{2}=0$ ($N_{2}=N_{3}=0$), we can obtain the maximum mean excitation number of the receiver mode,
\begin{widetext}
\begin{equation}
\langle a_{2}^{\dagger}a_{2}\rangle_{\text{max}}\!=\!\begin{dcases}
	\exp\left[-\frac{\kappa_{1(2)}}{\Lambda_{1}}\arctan\left(\frac{\kappa_{1(2)}\Lambda_{1}}{\kappa_{1(2)}^{2}-8g^{2}}\right)-\frac{\kappa_{1(2)}\pi}{\Lambda_{1}}\right], & 0\!\leq\!\kappa_{1(2)}\!\leq\!2\sqrt{2}g,\\
	\exp\left[-\frac{\kappa_{1(2)}}{\Lambda_{1}}\arctan\left(\frac{\kappa_{1(2)}\Lambda_{1}}{\kappa_{1(2)}^{2}-8g^{2}}\right)\right], & 2\sqrt{2}g\!<\!\kappa_{1(2)}\!<\!4g,\\
	e^{-2}, & \kappa_{1(2)}\!=\!4g,\\
	\exp\left[-\frac{\kappa_{1(2)}}{\sqrt{\kappa_{1(2)}^{2}-16g^{2}}}\ln\left(\frac{\kappa_{1(2)}+\sqrt{\kappa_{1(2)}^{2}-16g^{2}}}{\kappa_{1(2)}-\sqrt{\kappa_{1(2)}^{2}-16g^{2}}}\right)\right], & \kappa_{1(2)}\!>\!4g,
\end{dcases}~\label{db}
\end{equation}
\end{widetext}
with $\Lambda_{1}^{2}=16g^{2}-\kappa_{1(2)}^{2}>0$. The different branches in Eq.~(\ref{db}) correspond to different damping regimes of the effective two-mode dynamics, with $\kappa_{1(2)}=2\sqrt{2}g$ and $\kappa_{1(2)}=4g$ marking the critical points. To analyze the transfer of a single excitation in the two cases, we plot $\langle \hat{a}_{2}^{\dagger}\hat{a}_{2}\rangle_{\text{max}}$ as a function of the effective decay rates $\kappa_{1,2}$ when the single excitation is initially in the sender mode $a_{1}$, as shown in Fig.~\ref{N1_N2_N3_inf_v1}(b). We find that the value of  $\langle \hat{a}_{2}^{\dagger}\hat{a}_{2}\rangle_{\text{max}}$ decreases exponentially as $\kappa_{1,2}$ increases. The result means that $N_{1}$ ($N_{3}$) normal modes, which are only coupled with the sender (receiver), degrade the transfer efficiency as the effective coupling strength $G_{1}$ ($G_{2}$) increases. Note that the effective bath coupled only to the sender has the same influence on the transfer efficiency as the bath coupled only to the receiver. 

When the sender and the receiver are only coupled to the common bath, we can obtain the dynamical evolution of $\langle \hat{a}_{2}^{\dagger}\hat{a}_{2}(t)\rangle$ as
\begin{eqnarray}
	\langle \hat{a}_{2}^{\dagger}\hat{a}_{2}(t)\rangle & =&  -\frac{2(\gamma_{1}\gamma_{2}+4g^{2})}{\Lambda_{3}^{2}}\exp\left[-\frac{1}{2}t(\gamma_{1}+\gamma_{2})\right]\notag\\ &&\times\left[\cosh\left(\frac{\sqrt{\Lambda_{2}^{2}-\Lambda_{3}^{2}}}{2\sqrt{2}}t\right)-\cosh\left(\frac{\sqrt{\Lambda_{2}^{2}+\Lambda_{3}^{2}}}{2\sqrt{2}}t\right)\right],~\label{a2daga2t}
\end{eqnarray}
with
\begin{subequations}
	\begin{align}
		\Lambda_{2} & =  [(\gamma_{1}+\gamma_{2})^{2}-16g^{2}]^{1/2},\\
		\Lambda_{3} & = (	\Lambda_{2}^{4}+256g^{2}\gamma_{1}\gamma_{2})^{1/4}.
	\end{align}
\end{subequations}
Based on Eq.~(\ref{a2daga2t}), we can obtain the maximum mean excitation number of the receiver for the case of $g=0$ as
\begin{equation}
	\langle \hat{a}_{2}^{\dagger}\hat{a}_{2}\rangle_{\text{max}}=\frac{\gamma_{1}\gamma_{2}}{(\gamma_{1}+\gamma_{2})^{2}}.
\end{equation}
We find that $\langle \hat{a}_{2}^{\dagger}\hat{a}_{2}\rangle_{\text{max}}$  only depends on the decay rates $\gamma_{1}$ and $\gamma_{2}$. We plot $\langle \hat{a}_{2}^{\dagger}\hat{a}_{2}\rangle_{\text{max}}$ as a function of the decay rates $\gamma_{1}$ and $\gamma_{2}$, as shown in Fig.~\ref{N1_N2_N3_inf_v1}(c). We see from Fig.~\ref{N1_N2_N3_inf_v1}(c) that the maximum mean excitation number decreases as $\vert\gamma_{1}-\gamma_{2}\vert$ increases when $\gamma_{1}\neq\gamma_{2}$.  For $\gamma_{1}=\gamma_{2}$, the value of $\langle \hat{a}_{2}^{\dagger}\hat{a}_{2}\rangle_{\text{max}}$ is always equal to 0.25. This is because the dark mode consisting of the sender and receiver modes is decoupled from the effective bath 2, and 25\% of the excitations in the receiver mode $a_{2}$ associated with this dark mode cannot be transferred to the effective bath 2. Note that the maximum mean excitation number is symmetric about $\gamma_{1}=\gamma_{2}$.

When $g\neq 0$, we display the dependence of the maximum mean excitation number on $\gamma_{1}$ and $\gamma_{2}$ based on Eq.~(\ref{a2daga2t}) in Fig.~\ref{N1_N2_N3_inf_v1}(d). We find that the mean excitation number of the receiver mode $a_{2}$ can reach 1 for $\gamma_{1}=\gamma_{2}=0$, which means that the transfer efficiency from the sender to the receiver only through the connection $g$ is equal to 1. However, when $\gamma_{1}\neq 0$ and $\gamma_{2}\neq 0$, the transfer efficiency decreases as $\gamma_{1}$ ($\gamma_{2}$) increases at $\gamma_{1}>\gamma_{2}$ ($\gamma_{1}<\gamma_{2}$). Meanwhile, when $\gamma_{1}=\gamma_{2}$, the transfer efficiency first decreases and then approaches 0.25 as $\gamma$ increases, which can be explained with the dark-mode effect based on the dark-mode theorem~\cite{Jian2023}. 

We also investigate the single excitation transfer when either only one of $N_{1}$, $N_{2}$ and $N_{3}$ is 0, or none of them is 0. We plot the maximum mean excitation number $\langle \hat{a}_{2}^{\dagger}\hat{a}_{2}\rangle_{\text{max}}$ as a function of the effective decay rates $\{\kappa_{1,2},\gamma_{1,2}\}$ when the excitation is initially in the sender mode $a_{1}$. These results are shown in Figs.~\ref{N1_N2_N3_inf_v1}(e-k).  In Figs.~\ref{N1_N2_N3_inf_v1}(e) and~\ref{N1_N2_N3_inf_v1}(f), we show $\langle \hat{a}_{2}^{\dagger}\hat{a}_{2}\rangle_{\text{max}}$ as a function of $\kappa$ ($\kappa_{1}=\kappa$) and $ \gamma$ ($\gamma_{1}=\gamma_{2}=\gamma$) in the two cases ($\infty,\infty,0 |\text{No}$) and  ($\infty,\infty,0 |\text{Yes}$). We find that the phenomena are the same as those for the two cases of ($0,\infty,0 |\text{No}$) in Fig.~\ref{N1_N2_N3_inf_v1}(c) and ($0,\infty,0 |\text{Yes}$) in Fig.~\ref{N1_N2_N3_inf_v1}(d) for $\kappa=0$. When $\kappa\neq 0$, the value of $\langle \hat{a}_{2}^{\dagger}\hat{a}_{2}\rangle_{\text{max}}$ decreases to 0 as $\kappa$ increases, which means that the effective bath 1 inhibits the transfer of a single excitation from the sender to the receiver. When the sender and the receiver are coupled to two individual effective baths 1 and 3, marked ($\infty,0,\infty|\text{Yes}$), we show $\langle \hat{a}_{2}^{\dagger}\hat{a}_{2}\rangle_{\text{max}}$ as a function of $\kappa_{1}$ and $\kappa_{2}$ in Fig.~\ref{N1_N2_N3_inf_v1}(g). As either $\kappa_{1}$ or $\kappa_{2}$ increases, the value of $\langle \hat{a}_{2}^{\dagger}\hat{a}_{2}\rangle_{\text{max}}$ decreases from 1 to 0, which indicates that the individual effective baths degrade the transfer efficiency. In the two cases of ($0,\infty,\infty |\text{No}$) and  ($0,\infty,\infty |\text{Yes}$), we show $\langle \hat{a}_{2}^{\dagger}\hat{a}_{2}\rangle_{\text{max}}$ as a function of $\kappa$ ($\kappa_{2}=\kappa$) and $\gamma$ ($\gamma_{1}=\gamma_{2}=\gamma$) in Figs.~\ref{N1_N2_N3_inf_v1}(h) and~\ref{N1_N2_N3_inf_v1}(i). The phenomena are the same as those in the two cases of ($\infty,\infty,0 |\text{No}$) in Fig.~\ref{N1_N2_N3_inf_v1}(e) and  ($\infty,\infty,0 |\text{Yes}$) in Fig.~\ref{N1_N2_N3_inf_v1}(f). When all these three effective baths exist, we consider the cases $\kappa_{1,2}=\kappa$ and $\gamma_{1,2}=\gamma$, and show the maximum mean excitation number $\langle \hat{a}_{2}^{\dagger}\hat{a}_{2}\rangle_{\text{max}}$ of  the receiver as functions of $\kappa$ and $\gamma$ in Figs.~\ref{N1_N2_N3_inf_v1}(j) and \ref{N1_N2_N3_inf_v1}(k). We see that the value of $\langle \hat{a}_{2}^{\dagger}\hat{a}_{2}\rangle_{\text{max}}$ decreases to 0 as $\kappa$ increases, which means that the effective baths 1 and 3 suppress the excitation transfer. Based on the above analyses, we find that the effective vacuum bath 2 has a certain positive effect on the single-excitation transfer, while the effective vacuum baths 1 and 3 degrade the single-excitation transfer.

To include the dissipations, we assume that each node in the coupled bosonic network interacts with its own vacuum bath. Compared with the originally idealized configuration shown in Fig.~\ref{N1_N2_N3_inf_v1}(a), these environmental couplings will modify the system dynamics. For the sender mode $a_{1}$ and the receiver mode $a_{2}$, the presence of the baths implies the introduction of two additional decay channels, so their effective decay rates become $\tilde{\kappa}_{1,2} = \kappa_{1,2} + \kappa_{1,2}^{\prime}$, where $\kappa_{1}^{\prime}$ and $\kappa_{2}^{\prime}$ denote the decay rates of the sender mode $a_{1}$ and the receiver mode $a_{2}$ associated with their individually coupled environments, respectively. Similarly, the individual baths of these intermediate normal modes will lead to the modification of the decay rates $\gamma_{1,2}\rightarrow\tilde{\gamma}_{1,2}$.
In this case, the system can still be described within the open sender-receiver framework, and then the excitation transfer efficiency from modes $a_{1}$ to $a_{2}$ can also be described by Figs.~\ref{N1_N2_N3_inf_v1}(g), ~\ref{N1_N2_N3_inf_v1}(j), and ~\ref{N1_N2_N3_inf_v1}(k).

\section{Discussions and Conclusions~\label{DAD}}
Finally, we present some discussions on the experimental realization of the present scheme. In principle, our scheme can be realized based on all bosonic-mode platforms, provided that they possess three key capabilities: the controllable resonance frequencies of these bosonic modes, the tunable inter-mode couplings between these bosonic modes, and the preparation and measurement of the average excitation number in the sender and receiver modes. In our scheme, the bosonic mode can be realized with the superconducting transmission-line resonator, whose resonance frequency can be adjusted by controlling the magnetic flux threading through a series of SQUIDs embedded within the resonator~\cite{Castellanos2007}, where the resonance frequency varies from $2\pi\times 4$ GHz to $2\pi\times 7.8$ GHz. This allows the frequencies of these bosonic modes to be set according to the optimal transfer conditions identified in the bipartite-graph framework. Once the frequencies are tuned, the interaction strength between the modes can be independently controlled via a nonhysteretic rf-SQUID placed between the resonators, serving as a flux-tunable mutual inductance~\cite{Wulschner2016}. By varying the magnetic flux through the rf-SQUID, the inter-mode coupling can be tuned. In Ref.~\cite{Wulschner2016}, the nearly frequency-degenerate superconducting transmission line resonator have the resonance frequency $\omega_{1}/2\pi=6.482$ GHz, and the coupling strength $G/2\pi$ between the resonators varies from $-302$ MHz to $37$ MHz, named $G/\omega_{1}=-0.046\text{-}0.0057$. Thus, the parameters used in our simulations are within the reach of current experimental conditions.
With the frequencies and couplings tuned, the initial average excitation of the sender mode can be created by coupling the transmission-line resonator to a superconducting qubit. By driving the qubit with a proper pulse, the qubit can be excited. Then the single created excitation can be flipped into the transmission-line resonator by a $\pi/2$ resonant evolution~\cite{Houck2007}. In addition, the average excitation number of the receiver mode can be dispersively measured through the emission spectrum of the qubit. This is because the Stark shift of the qubit is related to $\langle a_{2}^{\dagger}a_{2}\rangle$ in the dispersive Jaynes-Cummings model~\cite{Houck2007}.
Based on the above discussions, we can see that these capabilities provide a practical route to the experimental implementation of our scheme.

In conclusion, we have proposed a bipartite-graph framework to regularize quantum excitation transfer in a (2+$N$)-mode coupled bosonic network consisting of one sender, one receiver, and an intermediate-mode subnetwork with $N$ bosonic modes. By diagonalizing the $N$ intermediate modes, the ($2+N$)-mode network is transformed into a bipartite-graph network, which can provide great advantage for analyzing quantum resource transfer on the network. In the bipartite-graph framework, we can simplify the network configuration and optimize the transfer paths, thereby obtain the parameter conditions for a highly efficient quantum transfer. We have also derived the quantum Langevin equations and established the relation between the covariance matrices in both the original and bipartite-graph representations. Furthermore, we have confirmed the excitation transfer efficiency in this network when a single excitation is initially in the sender.
For finite normal modes, we have found that the excitation transfer efficiency depends on both the network configurations and the system parameters. For infinite normal modes, these intermediate normal modes are treated as the effective vacuum baths containing discrete bosonic modes. We have derived quantum master equation to govern the evolution of the sender-receiver subnetwork under the Weisskopf-Wigner approximation, and revealed the relationship between the transfer efficiency and the effective decay rates for different coupling configurations.  It has been found that the effective vacuum bath coupled to either the sender or the receiver suppresses the transfer efficiency. In contrast, the common effective vacuum bath coupled to both the sender and the receiver provides a coherent channel that maintains a transfer efficiency of 0.25. The bipartite-graph framework will offer a new insight for understanding and implementing resource transfer in complex and large-scale classical and quantum networks.

\begin{acknowledgments}
	J.-Q.L. was supported in part by the National Natural Science Foundation of China (Grants No. 12575015, No. 12247105, and No. 12421005), National Key Research and Development Program of China (Grant No. 2024YFE0102400), and Hunan Provincial Major Sci-Tech Program (Grant No. 2023ZJ1010). L.-M.K. is supported by the NSFC (Grants No. 12247105, 11935006, 12175060 and 12421005), the Hunan Provincial Major Sci-Tech Program (Grant No. 2023ZJ1010), and the Henan Science and Technology Major Project (Grant No. 241100210400).
	F.N. is supported in part by: the Japan Science and Technology Agency (JST) [via the CREST Quantum Frontiers program Grant No. JPMJCR24I2, the Quantum Leap Flagship Program (Q-LEAP), and the Moonshot R\&D Grant No. JPMJMS2061]. C.L. was supported in part by Postgraduate Scientific Research Innovation Project of Hunan Province (Grant No. CX20250721) and Tenglong Innovative Talent Fund of Hunan Normal University (Grant No. 2025TL209). Y.-H.L. was supported by the Key Laboratory of
	Low Dimensional Quantum Structures and Quantum Control
	of Ministry of Education (Grant No. QSQC2508) and the
	Scientic Research Foundation of Education Bureau of Hunan
	Province (Grant No. 25B0857).
\end{acknowledgments}

\bibliography{ArticleRef}
\end{document}